\documentclass[preprint,tightenlines,aps,prd,showpacs,nofootinbib]{revtex4}
\usepackage{graphicx}
\usepackage{amsmath}
\usepackage{amssymb}
\usepackage{bm}

\begin{document}

\title{\mbox{}\\[10pt]
Line Shapes of the $\bm{X(3872)}$}
%%%%%%%%%%%%%%%%%%%%%%%%%%%%%%%%%%%%%%%%%%%%%%%%%%%%%%%%%%%%%%%%%%%%%%%%%%%%%%

\author{Eric Braaten and Meng Lu}
%\thanks{}
\affiliation{Physics Department, Ohio State University, Columbus,
Ohio 43210, USA}

\date{\today}
%%%%%%%%%%%%%%%%%%%%%%%%%%%%%%%%%%%%%%%%%%%%%%%%%%%%%%%%%%%%%%%%%%%%%%%%%%%%%%
\begin{abstract}
If the quantum numbers of the $X(3872)$ are $J^{PC}=1^{++}$, 
the measurement of its mass implies that it is either
a loosely-bound hadronic molecule whose
constituents are a superposition of the charm mesons pairs
$D^{*0} \bar D^0$ and $D^0 \bar D^{*0}$
or else it is a virtual state of these charm mesons.
Its binding energy is small enough that the decay width of a 
constituent $D^{*0}$ or $\bar D^{*0}$ has a significant effect
on the line shapes of the $X$ resonance.
We develop a simple approximation to the line shapes that takes
into account the effect of the $D^{*0}$ width as well as
inelastic scattering channels of the charm mesons.
We carry out a simultaneous fit to the line shapes in the 
$J/\psi \, \pi^+ \pi^-$ and $D^0 \bar D^0 \pi^0$ channels measured 
in the decays $B^+ \to K^+ + X$ by the Belle Collaboration.
The best fit corresponds to the $X(3872)$ being a bound state 
just below the $D^{*0} \bar D^0$ threshold, but a virtual state 
just above the $D^{*0} \bar D^0$ threshold is not excluded.
\end{abstract}

\pacs{12.38.-t, 12.39.St, 13.20.Gd, 14.40.Gx}
% 12.38.-t   Quantum chromodynamics
% 12.39.St  Factorization
% 13.20.Gd  Decays of J/psi, Upsilon, and other quarkonia
% 14.40.Gx   Mesons with S=C=B=0, mass > 2.5 GeV (including quarkonia)

%\keywords{}

\maketitle


\section{Introduction}

The $X(3872)$ is a hadronic resonance near 3872 MeV discovered in 2003
by the Belle Collaboration \cite{Choi:2003ue}
and subsequently confirmed by the CDF, Babar, and D0 Collaborations
\cite{Acosta:2003zx,Abazov:2004kp,Aubert:2004ns}.
In addition to the discovery decay mode $J/\psi \, \pi^+ \pi^-$, 
the $X$ has been observed to decay into $J/\psi \, \gamma$ and
$J/\psi \, \pi^+ \pi^-\pi^0$ \cite{Abe:2005ix}. 
The decay into $J/\psi \, \gamma$ implies that the $X$
is even under charge conjugation.  An analysis by the Belle Collaboration
of the decays of $X$ into $J/\psi \, \pi^+ \pi^-$ 
strongly favors the quantum numbers $J^{PC} = 1^{++}$,
but does not exclude $2^{++}$ \cite{Abe:2005iy}.
An analysis by the CDF Collaboration of the decays of 
$X$ into $J/\psi \, \pi^+ \pi^-$ is 
compatible with the Belle constraints \cite{Abulencia:2005zc}.
The Belle Collaboration 
has also discovered a near-threshold enhancement
in the $D^0 \bar D^0 \pi^0$ system near 3875 MeV \cite{Gokhroo:2006bt}.  
If this enhancement is associated with the $X(3872)$, the tiny phase space 
available would rule out $J=2$, leaving $1^{++}$ as the only option.

An important feature of the $X(3872)$ is that its mass is 
extremely close to the $D^{*0} \bar D^0$ threshold.  
The PDG value for $M_X$ comes from combining 
measurements of $X$ in the $J/\psi \, \pi^+ \pi^-$ decay mode
\cite{Yao:2006px}.  After taking
into account a recent precision measurement of the $D^0$ mass by the
CLEO Collaboration \cite{Cawlfield:2007dw}, the difference between
the PDG value for $M_X$ and the $D^{*0} \bar D^0$
threshold is 
%-----------------
\begin{equation}
M_X - (M_{*0} + M_0) = -0.6 \pm 0.6~{\rm MeV} , 
\label{MX-CLEO}
\end{equation}
%-----------------
where $M_{*0}$ and $M_0$ are the masses of $D^{*0}$ and $D^0$.
The negative central value is compatible with the $X$ being a bound
state of the charm mesons.
The PDG value for the mass of $X(3872)$ comes from combining 
measurements of $X$ in the $J/\psi \, \pi^+ \pi^-$ decay mode
\cite{Yao:2006px}.
The peak of the near-threshold enhancement in $D^0 \bar D^0 \pi^0$
 \cite{Gokhroo:2006bt} is at a mass satisfying
%-----------------
\begin{equation}
M - (M_{*0} + M_0) = +4.1 \pm 0.7^{+0.3}_{-1.6}~{\rm MeV} . 
\label{MX-DDbarpi}
\end{equation}
%-----------------
We have obtained this result from the value of $M$ quoted in 
Ref.~\cite{Gokhroo:2006bt} by subtracting $2 M_0 + (M_{*0} - M_0)$,
where $M_0$ is the 2006 PDG fitted mass for the $D^0$, 
and by dropping the error bar associated with the $D^0$ mass.
The positive central value in Eq.~(\ref{MX-DDbarpi}) 
is compatible with $X$ being a virtual state of charm mesons.
The difference between the masses in 
Eqs.~(\ref{MX-DDbarpi}) and (\ref{MX-CLEO})
is $4.7^{+1.0}_{-1.8}$ MeV.  This is more than two
standard deviations, which raises the question of whether 
the decays into $J/\psi \, \pi^+ \pi^-$ and  
$D^0 \bar D^0 \pi^0$ are coming from the same resonance.

The proximity of the mass of the $X(3872)$ to the
$D^{*0} \bar D^0$ threshold has motivated its identification as a
weakly-bound molecule whose constituents are a superposition
of the charm meson pairs $D^{*0} \bar D^0$ and $D^0 \bar{D}^{*0}$
\cite{Tornqvist:2004qy,Close:2003sg,Pakvasa:2003ea,Voloshin:2003nt}.
The establishment of the quantum numbers of the $X(3872)$
as $1^{++}$ would make this conclusion almost unavoidable.  
The reason is that these
quantum numbers allow S-wave couplings of the $X$ to
$D^{*0} \bar D^0$ and $D^0 \bar{D}^{*0}$.  
Nonrelativistic quantum mechanics implies that a resonance 
in an S-wave channel near a 2-particle threshold has special 
universal features \cite{Braaten:2003he}.  Because of the small 
energy gap between the resonance and the 2-particle threshold, 
there is a strong coupling between the resonance 
and the two particles.  This strong coupling 
generates dynamically a large length scale that can be 
identified with the absolute value of
the S-wave scattering length $a$ of the two particles.
Independent of the original mechanism for the resonance, 
the strong coupling transforms the resonance into a 
bound state just below the two-particle threshold if $a>0$ 
or into a virtual state just above the two-particle threshold if $a<0$.
If $a>0$, the bound state has a molecular structure,
with the particles having a large mean separation of order $a$.  

The universality of few-body systems with a large scattering length
has many applications in atomic, nuclear, and particle physics
\cite{Braaten:2004rn}.
To see that these universal features are relevant to the $X(3872)$, 
we need only note that its binding energy is small compared to 
the natural energy scale associated with pion
exchange: $m_\pi^2 / (2M_{*00}) \approx 10$ MeV, 
where $M_{*00}$ is the reduced mass of the two constituents. 
The universal features of the $X(3872)$ were first exploited by
Voloshin to describe its decays into $D^0 \bar D^0 \pi^0$ and
$D^0 \bar D^0 \gamma$, which can proceed through decay of the
constituent $D^{*0}$  or $\bar{D}^{*0}$ \cite{Voloshin:2003nt}.
Universality has also been applied to the
production process $B \to KX$ \cite{Braaten:2004fk,Braaten:2004ai},
to the line shapes of the $X$ \cite{Braaten:2005jj}, and
to decays of $X$ into $J/\psi$ and pions \cite{Braaten:2005ai}.
These applications rely on factorization formulas that separate
the length scale $a$ from all the shorter distance scales of QCD
\cite{Braaten:2005jj}.  The factorization formulas can be derived
using the operator product expansion for a low-energy
effective field theory \cite{Braaten:2006sy}.

Other interpretations of the $X(3872)$ 
besides a charm meson molecule or a charm meson virtual state
have been proposed, including a P-wave charmonium state 
or a tetraquark state. (For a review, see Ref.~\cite{Swanson:2006st}.)
The discrepancy between the masses $M_X$ and $M$ in Eqs.~(\ref{MX-CLEO})
and (\ref{MX-DDbarpi}) has been interpreted 
as evidence that the $J/\psi \, \pi^+ \pi^-$ events and
the $D^0 \bar D^0 \pi^0$ events arise from decays of two 
distinct tetraquark states whose masses differ by about 5 MeV
\cite{Maiani:2007vr}.  If the charmonium or tetraquark 
models were extended to include the coupling of
the $X$ to $D^{*0} \bar D^0$ and $D^0 \bar{D}^{*0}$
scattering  states, quantum mechanics implies that the tuning 
of the binding energy to the threshold region would transform 
the state into a charm meson molecule or a virtual state of charm mesons. 
Any model of the $X(3872)$ that does not take into account its 
strong coupling to charm meson scattering states 
should not be taken seriously.  

We will assume in the remainder of this paper that the 
quantum numbers of the $X(3872)$ are $1^{++}$, so that it has an S-wave
coupling to $D^{*0} \bar D^0$ and $D^0 \bar{D}^{*0}$.  In this case,
the measured mass $M_X$ implies unambiguously that $X$ must be either a 
charm meson molecule or a virtual state of charm mesons.
The remaining challenge is to discriminate between these two possibilities.
The Belle Collaboration has set an upper bound on the 
width of the $X$: $\Gamma_X < 2.3$~{\rm MeV} 
at the 90\% confidence level \cite{Choi:2003ue}.
If the width of the $X$ is sufficiently small, there is
a clear qualitative difference in the line shapes of
$X$ between these two possibilities.
We first consider the $D^0 \bar D^0 \pi^0$ decay mode,
which has a contribution from the decay of a constituent $D^{*0}$.
If the $X$ was a charm meson molecule, its line shape
in $D^0 \bar D^0 \pi^0$
would consist of a Breit-Wigner resonance below the $D^{*0} \bar D^0$
threshold and a threshold enhancement above the $D^{*0} \bar D^0$
threshold.  If the $X$ was a virtual state, 
there would only be the threshold enhancement above the $D^{*0} \bar D^0$
threshold.  We next consider decay modes that have no contributions 
from the decay of a constituent $D^{*0}$, such as $J/\psi \, \pi^+ \pi^-$.
If the $X$ was a charm meson molecule, its line shape in such a 
decay mode would be a Breit-Wigner resonance below the 
$D^{*0} \bar D^0$ threshold.  If the $X$ was a virtual state, 
there would only be a cusp at the $D^{*0} \bar D^0$ threshold.
The possibility of interpreting the $ X(3872)$ as a cusp at the 
$D^{*0} \bar D^0$ threshold has been suggested by Bugg \cite{Bugg:2004rk}.
Increasing the width of the $X$ provides
additional smearing of the line shapes.  This makes the qualitative 
difference between the line shapes of a charm meson molecule
and a virtual state less dramatic.
To discriminate between these two possibilities therefore requires 
a quantitative analysis.

The proposal that $X(3872)$ is a charm meson virtual state
has received support from a recent analysis 
by Hanhart et al.~\cite{Hanhart:2007yq} of data on
$B^+ \to K^+ + J/\psi \, \pi^+ \pi^-$ and 
$B^+ \to K^+ + D^0 \bar D^0 \pi^0$ from the Belle and Babar 
Collaborations.  They concluded that the $D^0 \bar D^0 \pi^0$ 
threshold enhancement is compatible with the $X(3872)$ 
only if it is a virtual state.
One flaw in the analysis of Ref.~\cite{Hanhart:2007yq} is that
they did not take into account the width of the constituent $D^{*0}$.  
They identified the rate for $D^0 \bar D^0 \pi^0$ 
with the rate for $D^{*0} \bar D^0$ and $D^0 \bar{D}^{*0}$
multiplied by the 62\% branching fraction for 
$D^{*0} \to D^0 \pi^0$.  Thus they assumed that 
$D^0 \bar D^0 \pi^0$ could not be produced below the 
$D^{*0} \bar D^0$ threshold.  Since a bound state has energy 
below the threshold, this implies that $X$ 
cannot decay into $D^0 \bar D^0 \pi^0$ if it is a bound state.
This contradicts one of the universal features of an S-wave 
threshold resonance.  As the binding energy of the resonance 
decreases, the mean separation of the constituents 
grows increasingly large,
suppressing all decay modes except those from the decay 
of a constituent.

In this paper, we develop a simple approximation to the 
line shapes of the $X(3872)$ that takes
into account the $D^{*0}$ width.
In Sec.~\ref{sec:widths}, we analyze the decays of the
$D^*$ mesons and give a simple expression for the 
energy-dependent width of a virtual $D^{*0}$.
In Sec.~\ref{sec:DDscat}, we give an expression for the 
scattering amplitude for $D^{*0} \bar D^0$ that takes into 
account the width of the $D^{*0}$ as well as 
inelastic scattering channels for the charm mesons.
In Sec.~\ref{sec:lineshape}, we use that scattering amplitude 
together with the factorization methods of 
Refs.~\cite{Braaten:2005jj,Braaten:2006sy}
to derive the line shapes of $X(3872)$ in the decays 
$B^+ \to K^+ + X$.
In Sec.~\ref{sec:SDfactor}, we give an order-of-magnitude
estimate of the short-distance factor associated with the 
$B^+ \to K^+$ transition in the factorization formula.
In Sec.~\ref{sec:hanhart}, we discuss the analysis of 
Ref.~\cite{Hanhart:2007yq} and point out the flaw from neglecting 
the effects of the $D^{*0}$ width.
In Sec.~\ref{sec:analysis}, we carry out a new analysis of the 
data on $B^+ \to K^+ + J/\psi \, \pi^+ \pi^-$ and 
$B^+ \to K^+ + D^0 \bar D^0 \pi^0$ from the Belle 
Collaboration, taking into account the effects of the 
$D^{*0}$ width.  In Sec.~\ref{sec:disc}, we discuss 
ways in which our description for the line shapes
could be further improved.

%\newpage

\section{$\bm{D^*}$ Widths}
\label{sec:widths}

In this section, we present a quantitative analysis 
of the decay widths of the $D^*$ mesons.  There are particles 
with six different masses that enter into the analysis.
We therefore introduce concise notation for the masses
of the charm mesons and the pions.
We denote the masses of the spin-0 charm mesons
$D^0$ and $D^+$ by $M_0$ and $M_1$, respectively.
We denote the masses of the spin-1 charm mesons
$D^{*0}$ and $D^{*+}$ by $M_{*0}$ and $M_{*1}$, respectively.
We denote the masses of the pions
$\pi^0$ and $\pi^+$ by $m_0$ and $m_1$, respectively.
In each case, the numerical subscript is the absolute value of the
electric charge of the meson.
The pion mass scale corresponding to either $m_0$ or $m_1$ 
will be denoted by $m_\pi$.
The result of a recent precision measurement of the $D^0$ mass by the
CLEO Collaboration is $M_0 = 1864.85 \pm 0.18$ MeV,
where we have combined the errors in quadrature \cite{Cawlfield:2007dw}.
We use the PDG values for the differences between the 
charm meson masses \cite{Yao:2006px}.
The errors on the pion masses are negligible compared to those
on the charm meson masses.
It is also convenient to introduce concise notations for simple
combinations of the masses.
We denote the reduced mass of a spin-1 charm meson
and a spin-0 charm meson by
%-----------------
\begin{eqnarray}
M_{*ij} &=& \frac{M_{*i}M_j}{M_{*i}+M_j} .
\end{eqnarray}
%-----------------
We denote the reduced mass of a pion
and a spin-0 charm meson by
%-----------------
\begin{eqnarray}
m_{ij} &=& \frac{m_i M_j}{m_i+M_j} .
\end{eqnarray}
%-----------------

The phase space available for the decay $D^* \to D \pi$ 
depends sensitively on the difference
between the $D^*$ mass and the $D \pi$ thresholds:
%-----------------
\begin{equation}
\delta_{ijk} = M_{*i} - M_j - m_k  .
\end{equation}
%-----------------
The differences between the $D^*$ masses and the thresholds
for $D \pi$ states with the same electric charge are
%-----------------
\begin{subequations}
\begin{eqnarray}
\delta_{000}
&=&  7.14 \pm 0.07~{\rm MeV} ,
\label{delta000}
\\
\delta_{011}
&=&  -2.23 \pm 0.12~{\rm MeV} ,
\label{delta011}
\\
\delta_{110}
&=&  5.66 \pm 0.10~{\rm MeV}  ,
\label{delta110}
\\
\delta_{101}
&=&  5.85 \pm 0.01~{\rm MeV}  .
\label{delta101}
\end{eqnarray}
\end{subequations}
%-----------------
The T-matrix elements for the decays  $D^* \to D \pi$ are proportional 
to the dot product of the polarization vector of the $D^*$
and the 3-momentum of the $\pi$. The coefficient of the dot product
can be expressed as the product of a constant,
which we denote by $\sqrt{3/2}\,  g/f_\pi$,
and a Clebsch-Gordan coefficient.
The partial widths for the decays $D^*\to D \pi$ are
%-----------------
\begin{subequations}
\begin{eqnarray}
\Gamma[D^{*+} \to D^0 \pi^+]
&=& \frac { 2 \sqrt{2} g^2 } {3 \pi f_\pi^2 }
m_{10}^{5/2} \delta_{101}^{3/2} ,
\label{GamD*+0+}
\\
\Gamma[D^{*+} \to D^+ \pi^0]
&=& \frac { \sqrt{2}  g^2 } {3 \pi f_\pi^2 }
m_{01}^{5/2} \delta_{110}^{3/2} ,
\label{GamD*++0}
\\
\Gamma[D^{*0} \to D^0 \pi^0]
&=& \frac { \sqrt{2} g^2 } {3 \pi f_\pi^2 }
m_{00}^{5/2} \delta_{000}^{3/2} .
\label{GamD*000}
\end{eqnarray}
\label{GamD*}
\end{subequations}
%-----------------
Since $\delta_{011}<0$, $D^{*0}$ does not decay into $D^+ \pi^-$.

The PDG value for the total width of the $D^{*+}$ is
$\Gamma[D^{*+}] = 96 \pm 22$ keV \cite{Yao:2006px}.
We can use the PDG values for the
branching fractions for $D^{*+}  \to  D^+ \pi^0$,
$D^{*+}  \to D^0 \pi^+$, and  $D^{*+}  \to  D^+ \gamma$
to determine the partial widths for those decays:
%-----------------
\begin{subequations}
\begin{eqnarray}
\Gamma[D^{*+} \to D^0 \pi^+] &=&  65.0 \pm 14.9 \  {\rm keV} ,
\label{GamD*+0+:N}
\\
\Gamma[D^{*+} \to D^+ \pi^0] &=&  29.5 \pm 6.8 \  {\rm keV} ,
\label{GamD*++0:N}
\\
\Gamma[D^{*+} \to D^+ \gamma\ ] &=& 1.5 \pm 0.5 \  {\rm keV} .
\label{GamD*++gam:N}
\end{eqnarray}
\end{subequations}
%-----------------
By fitting the partial widths in Eqs.~(\ref{GamD*+0+:N}) and
(\ref{GamD*++0:N})  to the expressions in Eqs.~(\ref{GamD*+0+})
and (\ref{GamD*++0}),
we obtain consistent determinations of the constant $g/ f_\pi$.
The more accurate of the two determinations  comes from the decay
$D^{*+} \to D^0 \pi^+$:
$g/ f_\pi = (2.82 \pm 0.32) \times 10^{-4}~{\rm MeV}^{-3/2}$.
Inserting this value into Eq.~(\ref{GamD*++0}), we can predict
the partial width for $D^{*0} \to D^0 \pi^0$:
%-----------------
\begin{eqnarray}
\Gamma[D^{*0} \to D^0 \pi^0] &=& 40.5 \pm 9.3 \  {\rm keV} .
\label{GamD*000:N}
\end{eqnarray}
%-----------------
By combining this prediction with the PDG value for the
branching fraction for $D^{*0} \to D^0 \pi^0$, we obtain
a prediction for the total width of the $D^{*0}$:
$\Gamma[D^{*0}] = 65.5 \pm 15.4$ keV.
We can then predict the
partial width for the decay $D^{*0} \to D^0 \gamma$
using the PDG value for its branching fraction:
%-----------------
\begin{eqnarray}
\Gamma[D^{*0} \to D^0 \gamma ] &=&  25.0 \pm 6.2 \  {\rm keV} .
\label{GamD*00gam}
\end{eqnarray}
%-----------------

%------------------------------------------------------------------------------------
\begin{figure}[t]
\includegraphics[width=12cm]{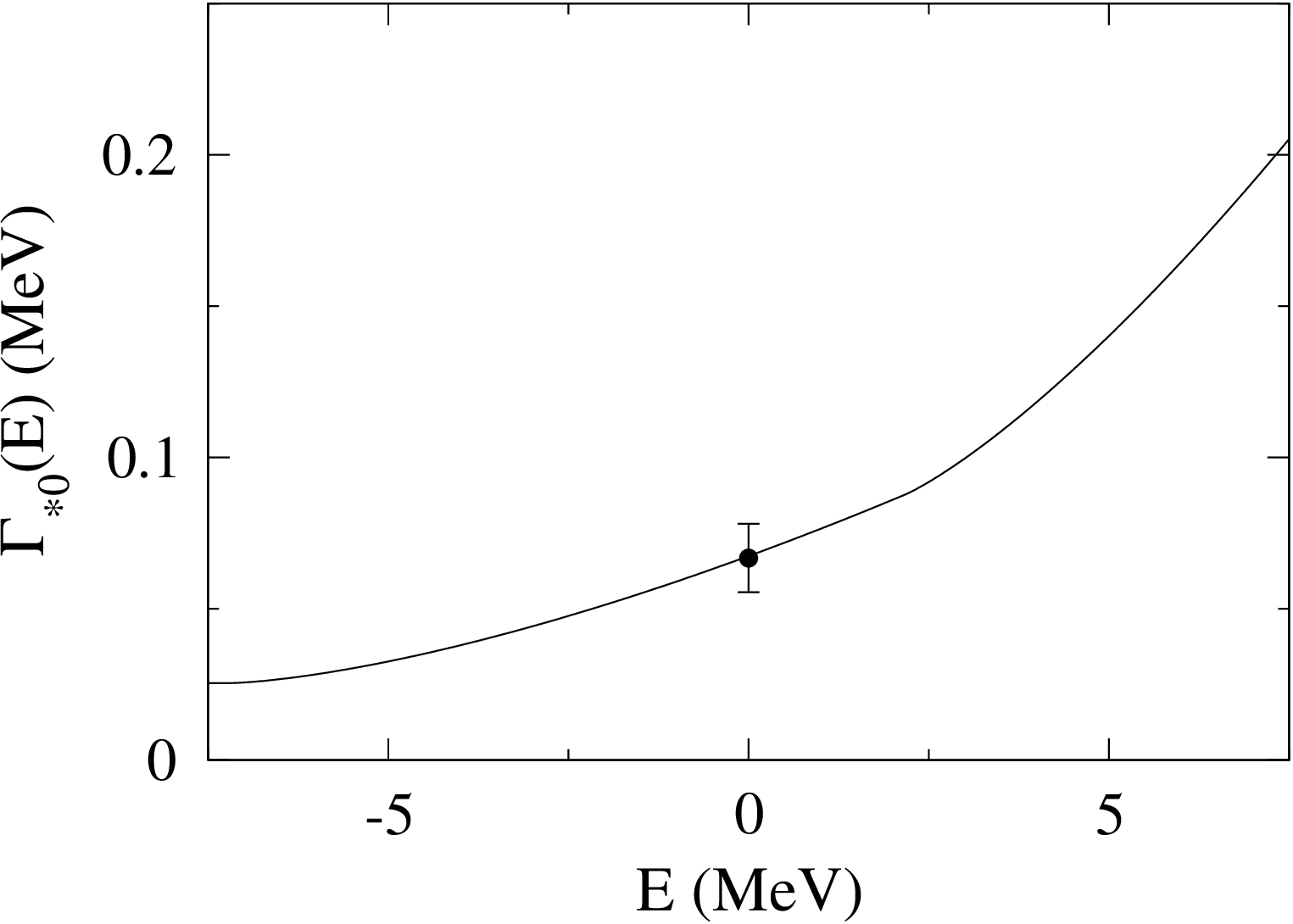}
\caption{
The energy-dependent width $\Gamma_{*0}(E)$ of a virtual $D^{*0}$ 
as a function of the energy $E$ relative to the $D^{*0}$ mass.
The point with error bars at $E=0$ indicates the value and 
uncertainty of the predicted width of $D^{*0}$.
\label{fig:Gam*}}
\end{figure}
%------------------------------------------------------------------------------------

Since the decay rates for $D^* \to D \pi$ in Eqs.~(\ref{GamD*})
scale like the 3/2 power of the
energy difference  between the $D^*$ mass and the $D \pi$ threshold,
they are fairly sensitive to the mass of the $D^*$.
A virtual $D^{*0}$
with energy $M_{*0}+E$ can be considered as a $D^{*0}$ whose
rest energy differs from its physical mass by the energy $E$.
The width of the virtual particle varies with $E$.
We denote the energy-dependent width of the $D^{*0}$
by $\Gamma_{*0}(E)$.
If $|E|$ is small compared to the pion mass, 
the energy-dependent width can be obtained simply by scaling 
the physical partial width for the decay $D^{*0}\to D^0 \pi^0$:
%-----------------
\begin{eqnarray}
\Gamma_{*0}(E) &=&
\Gamma[D^{*0} \to D^0 \gamma]
+ \Gamma[D^{*0} \to D^0 \pi^0]
\big[  \left[ (\delta_{000} + E)/\delta_{000} \right]^{3/2}
\theta(\delta_{000} + E)
\nonumber \\
&& \hspace{3cm}
+ 2 \left( m_{11}/m_{00} \right)^{5/2}
\left[ (\delta_{011} + E)/\delta_{000} \right]^{3/2}
\theta(\delta_{011} + E) \big].
\label{GamD*0-E}
\end{eqnarray}
%-----------------
We ignore any energy dependence of the decay widths into $D^0 \gamma$,
because the photon energy and the phase space
for the decays $D^{*0} \to D^0 \gamma$
do not vary dramatically in the $D^{*0} \bar D^0$ threshold region.
In Fig.~\ref{fig:Gam*}, we plot the energy-dependent width of the
$D^{*0}$ as a function of the energy $E$ relative to the
$D^{*0}$ mass. 
The three terms in Eq.~(\ref{GamD*0-E})
have obvious interpretations as energy-dependent partial widths
for decays of $D^{*0}$ into $D^{0} \gamma$, $D^{0} \pi^0$,
and $D^{+} \pi^-$.  We define  energy-dependent branching fractions
${\rm Br}_{00\gamma}(E)$, ${\rm Br}_{000}(E)$, and ${\rm Br}_{011}(E)$
by dividing these terms by $\Gamma_{*0}(E)$.
For example, the energy-dependent branching fraction
for $D^{*0} \to D^0 \pi^0$ is
%-----------------
\begin{equation}
{\rm Br}_{000}(E) =
\frac{\Gamma[D^{*0} \to D^0 \pi^0]}{\Gamma_{*0}(E)} \,
\left[ (\delta_{000} + E)/\delta_{000} \right]^{3/2}
\theta(\delta_{000} + E) .
\label{Br000}
\end{equation}
%-----------------

\section{Low-energy $\bm{D^{*0} {\bar D}^0}$ Scattering}
\label{sec:DDscat}

In this section, we discuss the low-energy scattering of the 
charm mesons $D^{*0}$ and $\bar D^0$.  The existence of the 
$X(3872)$ with quantum numbers $1^{++}$ implies that there 
is an S-wave resonance near threshold
in the channel with even charge conjugation:
%-----------------
\begin{eqnarray}
(D^* \bar D)^0_+ \equiv  \mbox{$\frac{1}{\sqrt{2}}$}
\left( D^{*0} \bar D^0  + D^0 \bar D^{*0} \right) .
\label{D*Dbar0}
\end{eqnarray}
%-----------------
The isospin splittings between the charm meson masses are
approximately 4.8 MeV for
$D^+ - D^0$ and 3.3 MeV for $D^{*+} - D^{*0}$.
The energy splitting between the $D^{*+} D^{-}$ and
$D^{*0} \bar D^0$ thresholds is the sum of these isospin splittings:
%-----------------
\begin{eqnarray}
\nu = 8.08 \pm 0.12~{\rm MeV}  .
\label{nu}
\end{eqnarray}
%-----------------
We will use approximations that are valid when the
energy $E$ relative to the $D^{*0} \bar D^0$
threshold is small compared to $\nu$. Thus the relative momenta 
of the charm mesons is required to be small compared to
$(2 M_{*00} \nu)^{1/2} \approx 125$ MeV.
This is numerically comparable to the pion mass scale:
$m_\pi \approx 135$ MeV.

The presence of the $X(3872)$ resonance so close to the
$D^{*0} \bar D^0$ threshold with quantum numbers that allow
S-wave couplings to $D^{*0} \bar D^0$ and $D^0 \bar D^{*0}$
indicates that it is necessary to treat the interaction
that produces the resonance nonperturbatively in
order to take into account the constraints of unitarity.
The resonance is in the channel $(D^* \bar D)^0_+$.  
There may also be scattering in the channel 
that is odd under charge conjugation, 
but we will assume that it can be neglected compared to  
scattering in the resonant channel.  Since the kinetic energy is so low,
the scattering will be predominantly S-wave. 

We express the transition amplitude $\mathcal{A}(E)$ 
for the scattering of nonrelativistically normalized
charm mesons in the channel $(D^* \bar D)^0_+$ in the form
%-----------------
\begin{equation}
\mathcal{A}(E) = \frac{2 \pi}{M_{*00}} f(E) ,
\end{equation}
%-----------------
where $f(E)$ is the conventional nonrelativistic scattering amplitude.
An expression for the low-energy scattering amplitude 
with an S-wave threshold resonance that is 
compatible with unitarity is 
%-----------------
\begin{equation}
f(E)  = \frac{1}{- \gamma + \kappa(E)} ,
\label{A0-ren}
\end{equation}
%-----------------
where $\kappa(E)= (- 2 M_{*00} E - i \varepsilon)^{1/2}$
and $E$ is the total energy of the charm mesons
relative to the $D^{*0} \bar D^0$ threshold.
If $E$ is real, the variable $\kappa(E)$ is real and positive 
for $E < 0$ and it is pure imaginary 
with a negative imaginary part for $E > 0$.
The parameter $\gamma$ in Eq.~(\ref{A0-ren})
can be identified as the inverse scattering length.
If $\gamma$ is complex, the imaginary part of the scattering amplitude 
in Eq.~(\ref{A0-ren}) satisfies
%-----------------
\begin{equation}
{\rm Im} \, f(E)  =
| f(E) |^2 \,
{\rm Im} \left[ \gamma - \kappa(E)  \right] .
\label{ImA-optical2}
\end{equation}
%-----------------

The scattering amplitude $f(E)$ in Eq.~(\ref{A0-ren}) 
satisfies the constraints of unitarity for a single-channel system 
exactly provided $\gamma$ is a real analytic function of $E$. 
For positive real values of the energy $E$,
Eq.~(\ref{ImA-optical2})
is simply the optical theorem for this single-channel system:
%-----------------
\begin{equation}
{\rm Im} \, f(E)  =
|  f(E) |^2
\sqrt{2 M_{*00} E}
\hspace{1cm} (E>0).
\label{ImA-optical:1ch}
\end{equation}
%-----------------
The left side is the imaginary part of the T-matrix element
for elastic scattering in the $(D^* \bar D)^0_+$ channel.
The right side is the cross section for elastic scattering
multiplied by $[E/(2 M_{*00})]^{1/2}$.
If $\gamma > 0$, the amplitude $f(E)$
has a pole at a negative value of the energy $E$,
indicating the existence of a stable bound state.
If $\gamma$ varies sufficiently slowly with $E$, 
we can approximate it by a constant.
The pole is then at $E_{\rm pole} \approx - \gamma^2/(2 M_{*00})$
and the binding energy is $\gamma^2/(2 M_{*00})$. 
In addition to the contribution to the imaginary part of $f(E)$
in Eq.~(\ref{ImA-optical:1ch}), there is a delta-function 
contribution at $E= E_{\rm pole}$:
%-----------------
\begin{equation}
{\rm Im} \, f(E)  \approx
\frac{\pi \gamma}{M_{*00}} 
\delta ( E + \gamma^2/(2 M_{*00})) 
\hspace{1cm} (E<0,\gamma>0).
\label{ImA-optical:1chE<0}
\end{equation}
%-----------------
If $\gamma < 0$,
the pole in the amplitude $f(E)$ is not on the real
$E$ axis, but on the second sheet of the complex variable $E$.
The standard terminology for such a pole is a {\it virtual state}.
The imaginary part of the amplitude is nonzero 
only in the positive $E$ region and is given by 
Eq.~(\ref{ImA-optical:1ch}).

We would like to identify the bound state in the case $\gamma > 0$
with the $X(3872)$.  However,
since the $X$ decays, it must have a nonzero width.
Its decay modes include $D^0 \bar D^0 \pi^0$ and
$D^0 \bar D^0 \gamma$, which receive contributions from decays
of the constituent $D^{*0}$ or $\bar D^{*0}$.  There are also
other decay modes, including $J/\psi \, \pi^+ \pi^-$,
$J/\psi \, \pi^+ \pi^- \pi^0$, and $J/\psi \, \gamma$.
All the decay modes of $X$ are inelastic scattering channels
for the charm mesons.  The dominant effects
of these inelastic scattering channels on scattering in the
$(D^* \bar D)^0_+$ channel can be taken into account through
simple modifications of the variables $\gamma$ and $\kappa(E)$
in the resonant amplitude $f(E)$ in Eq.~(\ref{A0-ren}).
The effects of the decays of the constituent $D^{*0}$
or $\bar{D}^{*0}$ can be taken into account simply by replacing
the mass $M_{*0}$ that is implicit in the energy $E$ measured
from the $D^{*0} \bar D^0$ threshold by
$M_{*0} - i \Gamma_{*0}(E)/2$,
where $\Gamma_{*0}(E)$ is the energy-dependent width
of the $D^{*0}$ given in Eq.~(\ref{GamD*0-E}).
This changes the energy variable $\kappa(E)$ 
defined after Eq.~(\ref{A0-ren}) to
%-----------------
\begin{equation}
\kappa(E)  = \sqrt{- 2 M_{*00} [E + i \Gamma_{*0}(E)/2]} .
\label{kappa-EGam}
\end{equation}
%-----------------
At the threshold $E=0$, the energy-dependent width $\Gamma_{*0}(E)$
reduces to the physical width $\Gamma[D^{*0}]$.
The expression for $\kappa(E)$ in Eq.~(\ref{kappa-EGam})
requires a choice of branch cut for the square root.
If $E$ is real, an explicit expression for $\kappa(E)$ 
that corresponds to the appropriate choice of branch cut
can be obtained by using the identity
%-----------------
\begin{eqnarray}
\sqrt{- 2 M [E + i \Gamma/2]}
&=&  \sqrt{M}
\left[ \left( \sqrt{E^2 + \Gamma^2/4} - E \right)^{1/2}
   - i  \left( \sqrt{E^2 + \Gamma^2/4} + E \right)^{1/2}
\right] .
\label{sqrtid}
\end{eqnarray}
%-----------------
The effects of inelastic channels other than $D^0 \bar D^0 \pi^0$ and
$D^0 \bar D^0 \gamma$ on scattering in the
$(D^* \bar D)^0_+$ channel can be taken into account by replacing
the real parameter $\gamma$ in Eq.~(\ref{A0-ren}) by a complex parameter
with a positive imaginary part.  With these modifications 
of the variables $\gamma$ and $\kappa(E)$,
the expression for the imaginary part of the amplitude  
$f(E)$ in Eq.~(\ref{ImA-optical2}) can now be interpreted 
as the optical theorem for this multi-channel system.
The right side can be interpreted as the total cross section 
for scattering in the $(D^* \bar D)^0_+$ channel
multiplied by $[E/(2 M_{*00})]^{1/2}$.
The term proportional to ${\rm Im} \, \gamma$ 
can be interpreted as the contribution from the inelastic 
scattering channels. This interpretation requires 
${\rm Im} \, \gamma > 0$.  The term proportional to 
$-{\rm Im} \, \kappa(E)$ includes a contribution
that reduces to the right side of Eq.~(\ref{ImA-optical:1ch})
in the limits ${\rm Im} \, \gamma \to 0$ and $\Gamma_{*0} \to 0$.
It can be interpreted as due to elastic scattering
into $D^{*0}\bar D^0$ or $D^0 \bar D^{*0}$.
If ${\rm Re} \, \gamma > 0$, the term proportional to 
$-{\rm Im} \, \kappa(E)$ also includes a contribution
that reduces to the right side of Eq.~(\ref{ImA-optical:1chE<0})
in the limits ${\rm Im} \, \gamma \to 0$ and $\Gamma_{*0} \to 0$. 
It can interpreted as due to scattering into the $X$ resonance. 
The $D^{*0}$ or $\bar D^{*0}$ produced by the elastic scattering process
will eventually  decay. Similarly, the constituent 
$D^{*0}$ or $\bar D^{*0}$ in the $X$ resonance will eventually decay.
Thus the ultimate final states corresponding to the term proportional 
to $-{\rm Im} \, \kappa(E)$
in Eq.~(\ref{ImA-optical2}) can be identified as
$D^0 \bar D^0 \pi^0$ and $D^0 \bar D^0 \gamma$, and also
$D^+ \bar D^0 \pi^-$ and $D^0 D^- \pi^+$ if the energy $E$ exceeds the 
threshold $|\delta_{011}| = 2.2$ MeV.

The scattering amplitude $f(E)$ given by Eqs.~(\ref{A0-ren})
and (\ref{kappa-EGam}) is a double-valued function 
of the complex energy $E$ with a square-root branch 
point and a pole. If we neglect the 
energy dependence of the width $\Gamma_{*0}(E)$,
the branch point is near $E = - i\Gamma[D^{*0}]/2$ and
the position of the pole is
%-----------------
\begin{equation}
E_{\rm pole}  \approx- \frac{\gamma^2}{2 M_{*00}} - \frac{i}{2} \Gamma[D^{*0}].
\label{Epole-gamma}
\end{equation}
%-----------------
If  ${\rm Re} \, \gamma > 0$,
the pole is on the physical sheet of the energy $E$.
It can be expressed in the form
%-----------------
\begin{equation}
E_{\rm pole}  =  - E_X - i \Gamma_X/2 ,
\label{Epole-EXGamX}
\end{equation}
%-----------------
where $E_X$ and $\Gamma_X$ are given by
%-----------------
\begin{subequations}
\begin{eqnarray}
E_X &\approx&  \left[ ({\rm Re} \, \gamma )^2 - ({\rm Im} \, \gamma)^2 \right]/(2 M_{*00}),
\\
\Gamma_X &\approx& \Gamma[D^{*0}]
+ 2 ({\rm Re} \, \gamma) ({\rm Im} \, \gamma)/M_{*00} .
\end{eqnarray}
\label{EXGamX}
\end{subequations}
%-----------------
If $E_X > 0$ and $\Gamma_X \ll E_X$, the state $X$
is a resonance whose line shape in the region $|E  - E_X| \ll E_X$
is a Breit-Wigner resonance centered at energy $-E_X$
with full width at half maximum $\Gamma_X$.
If $\Gamma_X$ is not small compared to $E_X$,
one can choose to define the binding energy and the width of $X$ 
to be given by the expressions for $E_X$ and $\Gamma_X$ in 
Eqs.~(\ref{EXGamX}), but they should not be interpreted literally.
If  ${\rm Re} \, \gamma < 0$, the pole at the energy $E_{\rm pole}$
in Eq.~(\ref{Epole-gamma}) is on the second sheet of the energy $E$
and it corresponds to a virtual state.
In this case, the expressions for $E_X$ and $\Gamma_X$ in 
Eqs.~(\ref{EXGamX}) have no simple physical interpretations.

%\newpage

\section{Line shapes of $\bm{X(3872)}$ in $\bm{B^+}$ decay}
\label{sec:lineshape}

If a set of particles $C$ has total quantum numbers that are compatible 
with those of the $X(3872)$ resonance and if the total energy $E$
of these particles is near the $D^{*0} \bar D^0$ threshold, 
then there will be a resonance in the channel $C$.
The line shape of $X(3872)$ in the channel $C$ is the differential rate 
as a function of the total energy $E$ of the particles in $C$.
In this section, we discuss the line shapes of the $X(3872)$
for energy $E$ close enough to the $D^{*0} \bar D^0$ threshold 
that we need only consider the resonant channel $(D^* \bar D)^0_+$
defined in Eq.~(\ref{D*Dbar0}).  This requires $|E|$ to be small 
compared to the energy $\nu \approx 8.1$ MeV
of the $D^{*+} D^-$ threshold. 

In Ref.~\cite{Braaten:2005jj}, it was pointed out that the line shape
for decay modes of $X$ that do not involve decays of a constituent
can be factored into short-distance factors
that are insensitive to $E$ and to the inverse scattering length $\gamma$
and a resonance factor that depends dramatically on $E$ and $\gamma$.
In Ref.~\cite{Braaten:2006sy}, it was shown that the factorization
formulas could be derived using the operator product expansion
for an effective field theory that describes the $c \bar c$
sector of QCD near the $D^{*0} \bar D^0$ threshold.
The factorization formulas hold for any {\it short-distance} process,
which we define to be one in which all the particles in the initial
state and all the particles in the final state beside the resonating
particles have momenta in the resonance rest frame that are of order
$m_\pi$ or larger.  An example of a short-distance process is the
discovery mode $B^+ \to K^+ + J/\psi \, \pi^+ \pi^-$.
In the $J/\psi \, \pi^+ \pi^-$ rest frame, the momenta of the
$B^+$ and $K^+$ are 1555 MeV, which is much larger than $m_\pi$.
The root-mean-square momentum
of the $J/\psi$ in the $J/\psi \, \pi^+ \pi^-$ rest frame is 319 MeV,
which is significantly larger than $m_\pi$.

We consider the short-distance process
$B^+ \to K^+ + C$ in which the collection of particles denoted by $C$
have a resonant enhancement associated with the $X(3872)$.
If the total energy $E$ of the particles in $C$
is near the $D^{*0} \bar D^0$ threshold,
the amplitude for the process
$B^+ \to K^+ + C$ factors into a short-distance
factor associated with the process
$B^+ \to K^+ + (D^* \bar D)^0_+$, a resonance factor $f(E)$ 
given by Eq.~(\ref{A0-ren}), and a short-distance factor  
associated with the process $(D^* \bar D)^0_+ \to C$.
Since the short-distance factors are insensitive to $E$, 
the only dramatic dependence on $E$ comes from the 
factor $f(E)$.  Thus the line
shape $d\Gamma/dE$ is proportional to $| f(E)|^2$.

For an alternative derivation of this line shape
that also gives the line shapes for 
$D^0 \bar D^0 \pi^0$ and $D^0 \bar D^0 \gamma$, 
we consider the inclusive differential decay rate into all 
channels that are enhanced by the $(D^* \bar D)^0_+$ resonance.
Using cutting rules, the inclusive decay rate
in the region near the $D^{*0} \bar D^0$ threshold 
summed over all resonant channels
can be expressed in the form
%-----------------
\begin{equation}
\frac{d\Gamma}{dE}[ B^+ \to K^+ + {\rm resonant}] =
2 \, \Gamma_{B^+}^{K^+} \, {\rm Im} \, f(E) ,
\label{lsh0:res}
\end{equation}
%-----------------
where $\Gamma_{B^+}^{K^+}$ is a short-distance factor 
defined in Ref.~\cite{Braaten:2006sy}.
The optical theorem in Eq.~(\ref{ImA-optical2}) can be used to
resolve this inclusive resonant rate into two terms.
We interpret the term proportional to ${\rm Im} \, \gamma$
as the sum of all contributions from short-distance channels.
Thus the line shape of $X(3872)$ in a specific short-distance 
channel $C$ can be expressed as
%-----------------
\begin{equation}
\frac{d\Gamma}{dE}[ B^+ \to K^+ + C] =
2 \, \Gamma_{B^+}^{K^+} \, |f(E)|^2 \, \Gamma^C(E).
\label{lsh0:SD}
\end{equation}
%-----------------
where $\Gamma_{B^+}^{K^+}$ is the same short-distance factor 
as in Eq.~(\ref{lsh0:res}) and $\Gamma^C(E)$ is  
a short-distance factor associated with the transition 
of the charm mesons to the particles in $C$.
The factor $\Gamma^C(E)$ in Eq.~(\ref{lsh0:SD}) 
differs by a factor of $\pi$ from the factor $\Gamma^C$
defined in Ref.~\cite{Braaten:2006sy}.
With this convention, $\Gamma^C(E)$ 
is the contribution of the state $C$ to the term
${\rm Im} \, \gamma$ in Eq.~(\ref{ImA-optical2}).
As indicated by the argument $E$, we
have allowed for the possibility that the dependence of this term 
on $E$ is not negligible in the $D^{*0} \bar D^0$ threshold region.
The justification for Eq.~(\ref{lsh0:SD}) relies on first
order perturbation theory in the coupling to the channel $C$.
If that coupling is too large, the channels $(D^* \bar D)^0_+$
and $C$ would have to be treated as a two-channel resonating system.
In the case of the channel $J/\psi \, \pi^+ \pi^-$, first
order perturbation theory can be justified by the small 
branching ratio of the decay of $X$ into $J/\psi \, \pi^+ \pi^-$
relative to $D^0 \bar D^0 \pi^0$ measured by the 
Belle Collaboration \cite{Gokhroo:2006bt}.

We interpret the term proportional to $- {\rm Im} \, \kappa(E)$
in Eq.~(\ref{ImA-optical2})
as the sum of all contributions from channels that correspond to 
$D^{*0} \bar D^0$ or $D^0 \bar D^{*0}$ followed by the decay of 
the $D^{*0}$ or $\bar D^{*0}$.  We can further resolve this 
term into the contributions from the channels 
$D^0 \bar D^0 \pi^0$, $D^0 \bar D^0 \gamma$, $D^+ \bar D^0 \pi^-$,
and $D^0 D^- \pi^+$ by multiplying it by the
energy-dependent branching fractions ${\rm Br}_{000}(E)$, 
${\rm Br}_{00\gamma}(E)$, $\frac{1}{2} {\rm Br}_{011}(E)$, 
and $\frac{1}{2} {\rm Br}_{011}(E)$, which add up to 1.
A simple expression for $-{\rm Im} \, \kappa(E)$ can be obtained 
by using the identity in Eq.~(\ref{sqrtid}).
The resulting expression for the line shape of $X$ in the 
$D^0 \bar D^0 \pi^0$ channel is 
%-----------------
\begin{equation}
\frac{d\Gamma}{dE}[ B^+ \to K^+ + D^0 \bar D^0 \pi^0] =
2 \, \Gamma_{B^+}^{K^+} \, | f(E) |^2
\left[ M_{*00} \big( \sqrt{E^2 + \Gamma_{*0}(E)^2/4} + E \big) \right]^{1/2}
{\rm Br}_{000}(E) ,
\label{lsh0:DDpi2}
\end{equation}
%-----------------
where $\Gamma_{B^+}^{K^+}$ is the same short-distance factor as in 
Eq.~(\ref{lsh0:res}) and
${\rm Br}_{000}(E)$ is given in Eq.~(\ref{Br000}).

%------------------------------------------------------------------------------------
\begin{figure}[t]
\includegraphics[width=12cm ]{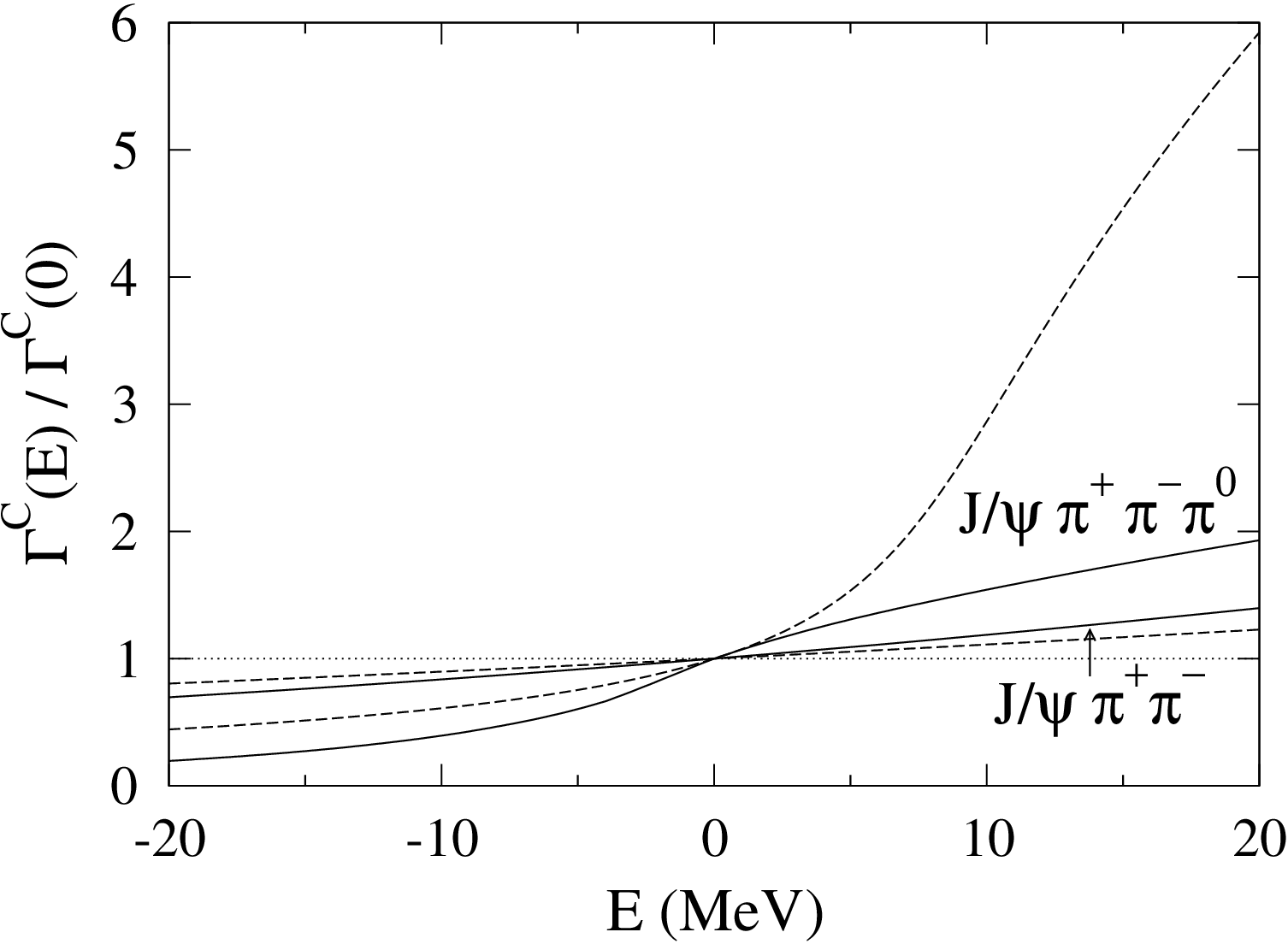}
\caption{
Energy dependence of the final-state factors 
$\Gamma^C(E)$ 
for the channels $C = J/\psi \, \pi^+ \pi^-$ 
and $J/\psi \, \pi^+ \pi^- \pi^0$.
The factors $\Gamma^{C}(E)/\Gamma^{C}(0)$ 
as functions of the energy $E$ relative to the
$D^{*0} \bar D^0$ threshold are shown as solid curves.
The dashed curves are the corresponding factors used in 
Ref.~\cite{Hanhart:2007yq}.
\label{fig:FC}}
\end{figure}
%-----------------------------------------------------------------------

In Ref.~\cite{Braaten:2005ai}, the decay rates of $X$ into $J/\psi$
plus pions and photons were calculated under the assumption that
these decays proceed through couplings of the $X$ to $J/\psi$
and the vector mesons $\rho^0$ and $\omega$.
The results of Ref.~\cite{Braaten:2005ai} can be used to calculate
the dependence of the factor $\Gamma^{C}(E)$ 
in Eq.~(\ref{lsh0:SD}) on the energy $E$
for $C= J/\psi \, \pi^+ \pi^-$, $J/\psi \, \pi^+ \pi^- \pi^0$,
$J/\psi \, \pi^0 \gamma$, and $J/\psi \, \gamma$.
The normalization of this factor, which we can take to be
$\Gamma^{C}(0)$, can only be determined by measurements 
of the branching fraction of $X(3872)$ into the final state $C$. 
In Fig.~\ref{fig:FC}, we plot the energy dependence of the
factors $\Gamma^{C}(E)$ for $C= J/\psi \, \pi^+ \pi^-$
and $J/\psi \, \pi^+ \pi^- \pi^0$. 
The larger variations for $J/\psi \, \pi^+ \pi^- \pi^0$
are due to the width of $\omega$ being much smaller than 
that of $\rho^0$.
Very close to the threshold,
the final state factors can be approximated by the expressions
%-----------------
\begin{subequations}
\begin{eqnarray}
\Gamma^{J/\psi \, \pi^+ \pi^-}(E) &\approx&  
\left( 1 + a_2 E \right) \Gamma^{\psi 2 \pi}_0 ,
\label{Gampsi2pi-E}
\\
\Gamma^{J/\psi \, \pi^+ \pi^- \pi^0}(E) &\approx&  
\left( 1 + a_3 E \right) \Gamma^{\psi 3 \pi}_0 ,
\label{Gampsi3pi-E}
\end{eqnarray}
\end{subequations}
%-----------------
where the coefficients are $a_2 = 0.0175 \, {\rm MeV}^{-1}$ and
$a_3 = 0.0809 \, {\rm MeV}^{-1}$. 
The approximations in Eqs.~(\ref{Gampsi2pi-E}) and (\ref{Gampsi3pi-E})
are accurate to within 1\% for 
$-8.5 \ {\rm MeV} < E < 10.1 \ {\rm MeV}$ 
and $-4.3 \ {\rm MeV} < E < 1.0 \ {\rm MeV}$, respectively.

In Fig.~\ref{fig:FC}, we also show the energy dependence of the
factors analogous to $\Gamma^{C}(E)$ that were used in 
Ref.~\cite{Hanhart:2007yq}.  Those factors were denoted 
by $\Gamma_{\pi^+ \pi^- J/\psi}(E)/g$
and $\Gamma_{\pi^+ \pi^- \pi^0 J/\psi}(E)/g$. 
In Ref.~\cite{Hanhart:2007yq}, the $\pi^+ \pi^-$ and 
$\pi^+ \pi^- \pi^0$ systems were treated simply as Breit-Wigner 
resonances with the masses and widths of the $\rho^0$ and $\omega$.
The additional dependence on $E$ from the coupling of the 
resonances to pions and from the integration over the phase space 
of the pions was not take into account.
In the region $|E| < 1$ MeV, the factors
in Ref.~\cite{Hanhart:2007yq} analogous to 
$\Gamma^{C}(E)/\Gamma^{C}(0)$ differ from ours by less than
1.4\%  for $C = J/\psi \, \pi^+ \pi^-$
and by less than 2.4\%  for $C= J/\psi \, \pi^+ \pi^- \pi^0$.
As illustrated in Fig.~\ref{fig:FC}, the differences are 
more substantial when $|E|$ is 10 MeV or larger
and they are particularly large for $J/\psi \, \pi^+ \pi^- \pi^0$.

%------------------------------------------------------------------------
\begin{figure}[t]
\includegraphics[width=12cm]{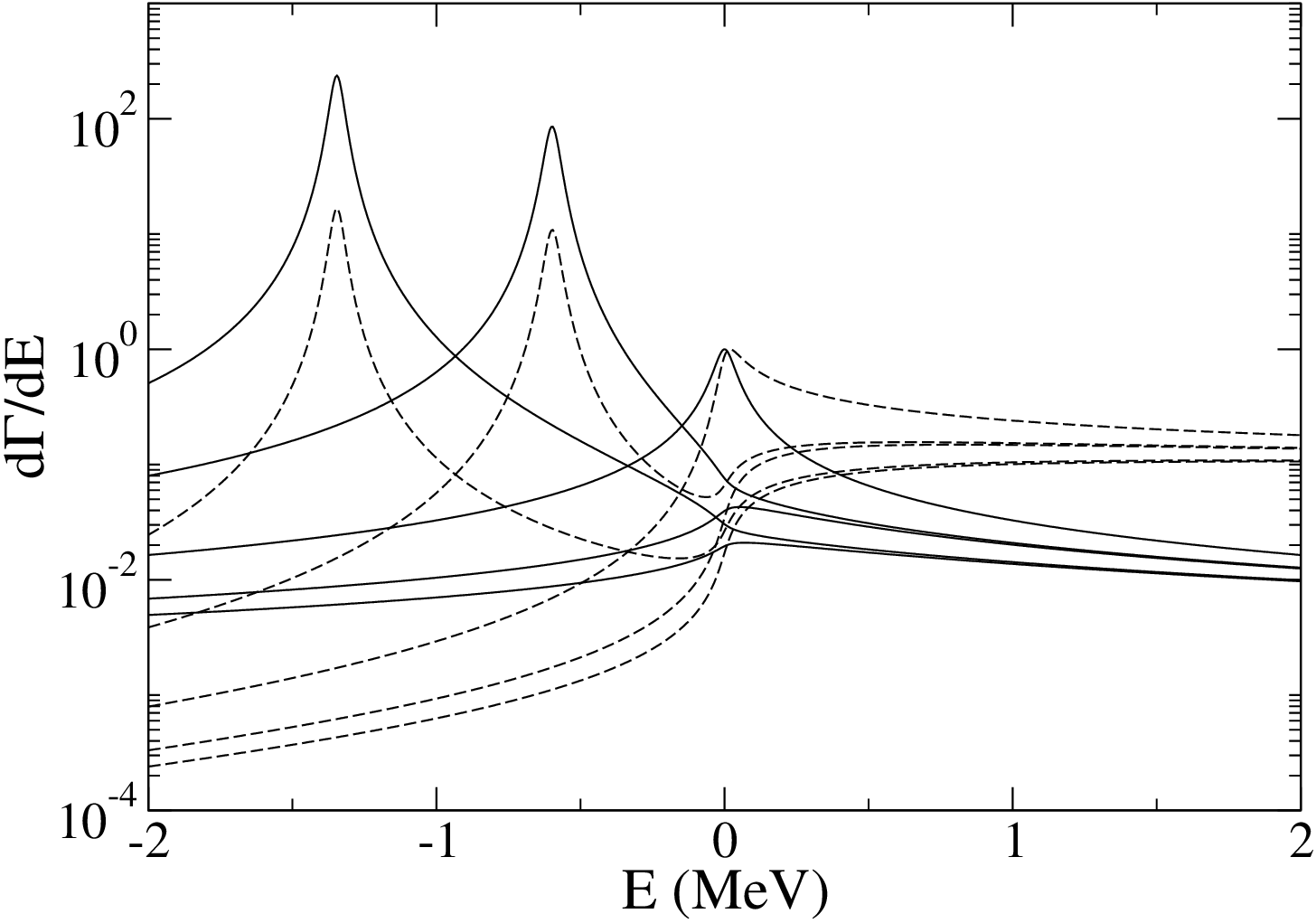}
\caption{The line shapes in a short-distance channel (solid lines)
and in the $D^0 \bar D^0 \pi^0$ channel (dotted lines).  
The line shapes are shown for five values of $\gamma$:
$-48$, $-34$, 0, 34, and 48 MeV.
The line shapes in the short-distance channel and in the 
$D^0 \bar D^0 \pi^0$ channel are separately normalized 
so that the curve for $\gamma = 0$ has a maximum value of 1.
At $E = 0.2$ MeV, the order of both the solid curves and 
the dashed curves from top to bottom is
$\gamma = 0$, 34, $-34$, 48, and $-48$ MeV.
\label{fig:LSsdd1} }
\end{figure}
%------------------------------------------------------------------------------------

In Fig.~\ref{fig:LSsdd1}, we illustrate the line shapes of $X(3872)$ 
in the $D^0 \bar D^0 \pi^0$ channel and in a short-distance channel, 
such as $J/\psi \, \pi^+ \pi^-$.  We take the parameter $\gamma$
to be real, which corresponds to the assumption that decay modes other than
$D^0 \bar D^0 \pi^0$ and $D^0 \bar D^0 \gamma$ give a small contribution
to the total width of $X(3872)$.
In Fig.~\ref{fig:LSsdd1}, 
the solid lines are the line shapes for a short-distance decay mode,
which is given by Eq.~(\ref{lsh0:SD}).  We neglect the energy-dependence 
from the factor $\Gamma^C(E)$, which is very small for $J/\psi \, \pi^+ \pi^-$.    
The dashed lines are the line shapes for $D^0 \bar D^0 \pi^0$,
which is given by Eq.~(\ref{lsh0:DDpi2}).
We show the line shapes for five values of $\gamma$:
48, 34, 0, $-34$, and $-48$ MeV.
For $\gamma = 34$, 0, and 48 MeV, the peaks of the resonance 
are at $E = -0.6$, 0, and $-1.2$ MeV, respectively. 
They correspond to the central value of the measurement 
in Eq.~(\ref{MX-CLEO}) and deviations by $\pm1\sigma$ 
from the central value.
In Fig.~\ref{fig:LSsdd1}, the line shapes for the short-distance channel
and for the $D^0 \bar D^0 \pi^0$ channel are separately normalized 
so that the maximum value is 1 for $\gamma = 0$.
This figure illustrates that as $\gamma$ decreases toward 0 from 
above, the area under the short-distance line shape decreases 
and it becomes very small for negative values of $\gamma$. 
As $\gamma$ decreases toward 0 from above, 
the area under the $D^0 \bar D^0 \pi^0$ line shape decreases less
rapidly, but it is also  small for negative values of $\gamma$.
At the values $\gamma = 48$, $34$, 0, $-34$, 
and $-48$ MeV, the differences between the positions of the peaks 
in the $D^0 \bar D^0 \pi^0$ channel and the short-distance channel
are 0.00024, 0.00052, 0.019, 0.75, and 1.75 MeV, respectively.
Thus the $4.7^{+1.0}_{-1.8}$ MeV difference between the masses in 
Eqs.~(\ref{MX-DDbarpi}) and (\ref{MX-CLEO}) appears to be only 
compatible with negative values of $\gamma$.

%------------------------------------------------------------------------
\begin{figure}[t]
\includegraphics[width=12cm]{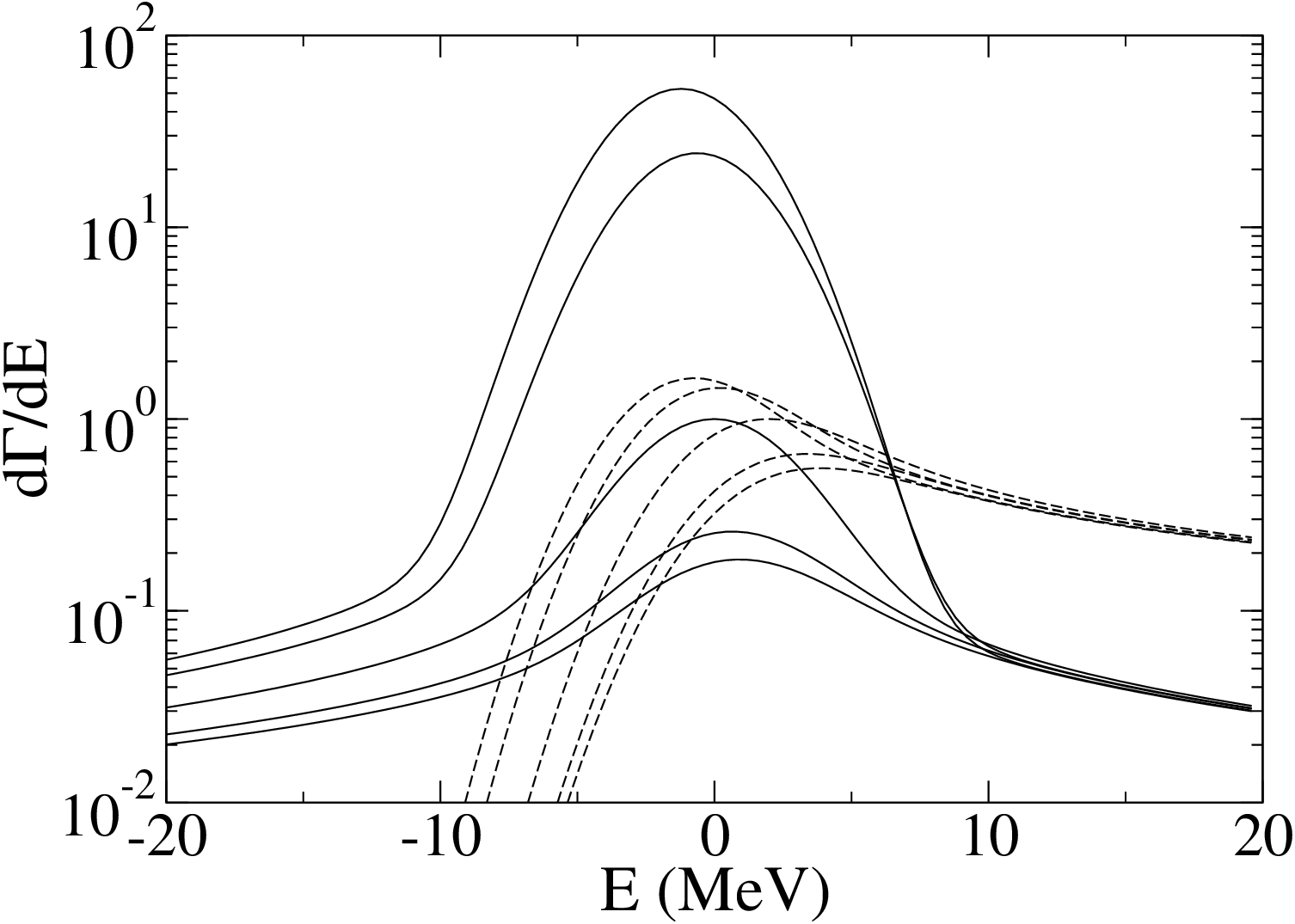}
\caption{
Same as in Fig.~\ref{fig:LSsdd1},
except that the line shapes have been smeared
by a Gaussian function with width 2.5 MeV
to mimic the effects of experimental resolution.
The smeared line shapes in the short-distance channel and in the 
$D^0 \bar D^0 \pi^0$ channel are separately normalized 
so that the curve for $\gamma = 0$ has a maximum value of 1.
At $E = 0$ MeV, the order of both the solid curves and 
the dashed curves from top to bottom is
$\gamma = 48$, 34, 0, $-34$, and $-48$ MeV.
\label{fig:LSsdd2} }
\end{figure}
%------------------------------------------------------------------------------------

In an actual measurement, the line shapes shown in 
Fig.~\ref{fig:LSsdd1} would be smeared by the effects of experimental
resolution.  In the Belle discovery paper, the $X(3872)$ signal 
in $J/\psi \, \pi^+ \pi^-$ was fit by a Gaussian with 
width 2.5 MeV, which is compatible with the experimental 
resolution \cite{Choi:2003ue}.  For the $D^0 \bar D^0 \pi^0$ 
enhancement near threshold, the signal was fit by a Gaussian with
width 2.24 MeV, which is again compatible with the experimental 
resolution \cite{Abe:2005ix}.  In Fig.~\ref{fig:LSsdd2},
we illustrate the effects of experimental resolution by 
smearing the line shapes in Fig.~\ref{fig:LSsdd1} by a Gaussian 
with width 2.5 MeV.  At the values $\gamma = 48$, $34$, 0, $-34$, 
and $-48$ MeV, the differences between the positions of the peaks 
in the $D^0 \bar D^0 \pi^0$ channel and the short-distance channel
are 0.45, 0.82, 2.0, 2.4, and 3.2 MeV, respectively.
Thus, after taking into account the effects of experimental 
resolution, the $4.7^{+1.0}_{-1.8}$ MeV difference 
between the masses in Eqs.~(\ref{MX-DDbarpi}) and (\ref{MX-CLEO})
is not incompatible with a positive value of $\gamma$.

\section{Estimate of the short-distance factor}
\label{sec:SDfactor}

In this section, we give an order-of-magnitude estimate of the 
short-distance factor $\Gamma_{B^+}^{K^+}$  that appears in 
Eqs.~(\ref{lsh0:res}), (\ref{lsh0:SD}), and (\ref{lsh0:DDpi2}).
We use the measurement by the Babar Collaboration of the 
branching fraction for the decay of $B^+$ into $K^+ D^{*0} \bar D^0$
\cite{Aubert:2003jq}:
%-----------------
\begin{equation}
{\rm Br}[B^+ \to K^+ D^{*0} \bar D^0] = 
( 4.7 \pm 1.0) \times 10^{-3}.
\label{BrBKD*D}
\end{equation}
%-----------------
The upper bound on the branching fraction for $K^+ D^0 \bar D^{*0}$
is $3.7 \times 10^{-3}$ at the 90\% confidence level.
The partial width for the decay in Eq.~(\ref{BrBKD*D}) can be written
%-----------------
\begin{equation}
\Gamma[B^+ \to K^+ D^{*0} \bar D^0] = 
\frac{3}{2 M_{B}} \left\langle | \mathcal{M} |^2 \right\rangle
\Phi_{B^+}^{K^+ D^{*0} \bar D^0} ,
\label{GamBKD*D}
\end{equation}
%-----------------
where $\Phi_{B^+}^{K^+ D^{*0} \bar D^0} = 194.5~{\rm MeV}^2$ 
is the integral of the three-body phase space
and $3 \langle | \mathcal{M} |^2 \rangle$ is the square of the 
matrix element averaged over the relativistic three-body phase space 
and summed over the $D^{*0}$ spins.
Using the measured lifetime, $\tau[B^+]= 1.64 \times 10^{-12}$ s, 
and the central value of the branching fraction in Eq.~(\ref{BrBKD*D}), 
we find that the average
value of the square of the matrix element in Eq.~(\ref{GamBKD*D})
is $\langle | \mathcal{M} |^2 \rangle = 3.4 \times 10^{-11}$.

The differential rate for $B^+$ to decay into 
$K^+ + D^{*0} \bar D^0$ with $D^{*0} \bar D^0$
near its threshold can be written  
%-----------------
\begin{equation}
\frac{d\Gamma}{dE}[B^+ \to K^+  + D^{*0} \bar D^0] = 
\frac{3\lambda^{1/2}(M_{B},M_{*0} + M_0,m_K)}{64 \pi^3 M_{B}^3} | \mathcal{M} |^2
\sqrt{2 M_{*00} E} ,
\label{dGamBKD*D}
\end{equation}
%-----------------
where $E$ is the rest energy of the $D^{*0} \bar D^0$ relative to 
its threshold and $| \mathcal{M} |^2$ is the square of the matrix element 
averaged over angles in the $D^{*0} \bar D^0$ rest frame
and averaged over the $D^{*0}$ spins.  The dramatic dependence of 
$| \mathcal{M} |^2$ on $E$ comes from the resonance factor 
$| f(E) |^2$ in Eq.~(\ref{lsh0:DDpi2}).
By comparing Eq.~(\ref{dGamBKD*D}) with  Eq.~(\ref{lsh0:DDpi2}), 
we can obtain an expression for the short-distance factor
$\Gamma_{B^+}^{K^+}$:
%-----------------
\begin{equation}
\Gamma_{B^+}^{K^+} = 
\frac{3\lambda^{1/2}(M_{B},M_{*0} + M_0,m_K)}{128 \pi^3 M_B^3}
\lim_{E \to 0}  |\mathcal{M}|^2/|f(E)|^2 .
\label{GamBKlim}
\end{equation}
%-----------------
 
Because of the threshold resonance, the square of the 
matrix element $| \mathcal{M} |^2$ for $B^+ \to K^+  + D^{*0} \bar D^0$
should be significantly larger near the $D^{*0} \bar D^0$ threshold
than its average $\langle | \mathcal{M} |^2 \rangle$ 
over the entire three-body phase.
We will assume that this threshold resonance is the only effect 
that gives a significant enhancement of $| \mathcal{M} |^2$ 
in this corner of the phase space.
Thus as an order of magnitude estimate of $| \mathcal{M} |^2$, 
we will take
%-----------------
\begin{equation}
| \mathcal{M} |^2 \approx
\left\langle | \mathcal{M} |^2 \right\rangle
 \frac{\Lambda^2}{|- \gamma + \kappa(E)|^2} ,
\label{Msqapprox}
\end{equation}
%-----------------
where $ \langle | \mathcal{M} |^2 \rangle$ is the average
value of the square of the matrix element defined by 
Eq.~(\ref{GamBKD*D}) and
$\Lambda$ is the momentum scale below which the 
resonant behavior sets in.  The natural scale for $\Lambda$ is 
$( 2 M_{*11} \nu)^{1/2}= 125$ MeV or $m_\pi= 135$ MeV.
The estimate in Eq.~(\ref{Msqapprox}) is likely to be an 
underestimate because $\langle | \mathcal{M} |^2 \rangle$
is enhanced by various resonant contributions,
including $D_{s1}(2536) +  \bar D^0$.
Using the value of $\langle | \mathcal{M} |^2 \rangle$
that corresponds to the central value of the branching fraction 
in Eq.~(\ref{BrBKD*D}), our estimate of the short-distance factor is 
%-----------------
\begin{equation}
\Gamma_{B^+}^{K^+} \approx \left( 3.8 \times 10^{-14}~{\rm MeV} \right)
\frac{\Lambda^2}{m_\pi^2} .
\label{GamBKest}
\end{equation}
%-----------------

\section{Analysis of Ref.~\cite{Hanhart:2007yq}}
\label{sec:hanhart}

In Ref.~\cite{Hanhart:2007yq}, the authors analyzed the data 
from the Belle and Babar
collaborations on $B \to K+ X(3872)$ using a generalized Flatt\'e
parameterization for the $(D^* \bar D)^0_+$ scattering amplitude:
\begin{equation}
f_{\rm HKKN}(E) = 
\left( - \frac 2 g \left[ E - E_f + i \Gamma_{\pi^+\pi^-J/\psi}(E)/2 
+ i \Gamma_{\pi^+\pi^-\pi^0J/\psi}(E)/2 \right] 
+ \kappa(E) + \kappa_1(E) \right)^{-1},
\label{f:HKKN}
\end{equation}
where $\kappa (E) = (-2 M_{*00} E + i \epsilon)^{1/2}$ and 
$\kappa_1 (E) = (-2 M_{*11} (E - \nu) + i\epsilon)^{1/2}$.
The functions 
$\Gamma_{\pi^+\pi^-J/\psi}(E)$ and 
$\Gamma_{\pi^+\pi^-\pi^0J/\psi}(E)$ are 
determined by the $\rho$ and $\omega$ resonance parameters
up to normalization factors $f_\rho$ and $f_\omega$. The resonant
term in the differential branching fraction for the inclusive decay 
of $B^+$ into $K^+$ was expressed in the form
\begin{equation}
\frac{d {\rm Br}}{d E} \left [ B^+ \to K^+ + {\rm resonant} \right ] =
\frac{2}{\pi} (\mathcal{B}/g) \, {\rm Im} \, f(E) ,
\label{BKres:HKKN}
\end{equation}
The adjustable parameters in the model of Ref.~\cite{Hanhart:2007yq} are $g$,
$E_f$, $f_\rho$, $f_\omega$, and ${\mathcal B}$. The imaginary part
of the scattering amplitude in Eq.~(\ref{f:HKKN}) is
\begin{eqnarray}
{\rm Im} \, f_{\rm HKKN}(E) &=& | f(E)|^2 
\left[ \Gamma_{\pi^+\pi^-J/\psi}(E)/g 
+ \Gamma_{\pi^+\pi^-\pi^0J/\psi}(E)/g 
- {\rm Im} \, \kappa(E) - {\rm Im} \, \kappa_1(E) \right] .
\nonumber
\\
\label{Imf:HKKN}
\end{eqnarray}
The first two terms in the square brackets are identified with the
contributions from $J/\psi \, \pi^+ \pi^-$
and $J/\psi \, \pi^+ \pi^- \pi^0$, respectively.  The third term
$- {\rm Im} \, \kappa(E)$
is nonzero only for $E > 0$ and was identified with the 
contribution from $D^0 \bar D^0 \pi^0$.  The last term
$- {\rm Im} \, \kappa_1(E)$
is nonzero only for $E > \nu$ and could be identified with the 
contributions from $D^{*+} D^-$ and $D^+ D^{*-}$. 

In the analysis of Ref.~\cite{Hanhart:2007yq},
the ratio $f_\omega / f_\rho$ was
determined from the branching ratio of $X(3872)$ into 
$J/\psi \, \pi^+ \pi^- \pi^0$ and $J/\psi \, \pi^+ \pi^-$ measured 
by the Belle collaboration \cite{Abe:2005ix}. 
The parameters $g$, $E_f$, $f_\rho$, and ${\mathcal B}$ were then
used to fit the $J/\psi \, \pi^+ \pi^-$ energy distributions in the decay
$B^+ \to K^+ + J/\psi \, \pi^+ \pi^-$ measured by the Belle and
Babar Collaborations  \cite{Choi:2003ue,Aubert:2005zh}
and the $D^0 \bar D^0 \pi^0$ energy distribution  in the decay
$B^+ \to K^+ + D^0 \bar D^0 \pi^0$ measured by the 
Belle collaboration \cite{Gokhroo:2006bt}.
The authors of Ref.~\cite{Hanhart:2007yq} found that their fits 
exhibited scaling behavior. The fits were insensitive to rescaling 
all the parameters $g$, $E_f$, $f_\rho$, $f_\omega$, and $\mathcal{B}$ 
by a common factor. From the expressions in Eqs.~(\ref{f:HKKN}), 
(\ref{BKres:HKKN}), and (\ref{Imf:HKKN}), one can see that the scaling 
behavior simply implies that the term $-(2/g)E$ in the denominator 
of Eq.~(\ref{f:HKKN}) can be neglected. 
The authors interpreted the parameter $g$ as a $D \bar D^*$
coupling constant. However a more appropriate interpretation of $g$ is
that it determines the effective range $r_s$ for S-wave scattering in the
$(D^* \bar D)^0_+$ channel. If we neglect the small imaginary terms
proportional to $\Gamma_{\pi^+\pi^-J/\psi}(E)$ and 
$\Gamma_{\pi^+\pi^-\pi^0J/\psi}(E)$ in the denominator in 
Eq.~(\ref{f:HKKN}), the inverse scattering length 
and the effective range are
\begin{subequations}
\begin{eqnarray}
\gamma & \approx & - 2 (E_f/g) - \sqrt{2 M_{*11} \nu} ,
\label{gamma:HKKN}
\\
r_s & \approx & -2/(M_{*00} \, g) - 1/\sqrt {2M_{*11} \nu} .
\label{rs:HKKN}
\end{eqnarray}
\end{subequations}
We have simplified the expression for $r_s$ in Eq.~(\ref{rs:HKKN})
by setting $M_{*11}/M_{*00}=1$.
The inverse scattering lengths $\gamma$ 
calculated from Eq.~(\ref{gamma:HKKN}) using the parameters of the 
fits in Ref.~\cite{Hanhart:2007yq}
range from $-51.7$ MeV to $-66.3$ MeV.
These values are close to the exact results for the 
complex inverse scattering length $\gamma$ for the fits in
Ref.~\cite{Hanhart:2007yq}.  Their real parts ranged 
from $-48.9$ MeV to $-63.5$ MeV
and their imaginary parts ranged from $3.3$ MeV to $6.8$ MeV.
Because of the scaling behavior,
the scattering amplitude near the $D^{*0} \bar D^0$ threshold
is insensitive to the effective range $r_s$, so its numerical value 
cannot be determined from the fits of Ref.~\cite{Hanhart:2007yq}. 
The authors could have obtained essentially 
the same fits by taking the range $r_s$ to be 0.  They could have 
replaced the scattering amplitude in Eq.~(\ref{f:HKKN}) by
\begin{equation}
f_{\rm HKKN}(E) \approx 
\left[ - {\rm Re} \, \gamma - i \Gamma_{\pi^+\pi^-J/\psi}(E)/g
- i \Gamma_{\pi^+\pi^-\pi^0J/\psi}(E) /g
+ \kappa (E) \right]^{-1} ,
\label{fit:scat-amp}
\end{equation}
with ${\rm Re} \, \gamma$, $f_\rho/g$, $f_\omega/g$, and
$\mathcal{B}/g$ as the adjustable parameters.
 
The authors of Ref.~\cite{Hanhart:2007yq} found that they could obtain 
acceptable fits to the Belle and Babar data only if the 
$J/\psi \, \pi^+ \pi^-$ energy
distribution was peaked almost exactly at the $D^{*0} \bar D^0$ 
threshold and the $D^0 \bar D^0 \pi^0$ energy distribution 
was peaked 2 to 3 MeV above the $D^{*0} \bar D^0$ threshold.
This requires a negative value for ${\rm Re} \, \gamma$,
which would correspond to a virtual state. 
They concluded that if the $D^0 \bar D^0 \pi^0$ threshold enhancement
is associated with the $X(3872)$, then $X$ must be a 
virtual state of charm mesons.

A crucial flaw in the analysis of Ref.~\cite{Hanhart:2007yq}
is that they assumed that the
$D^0 \bar D^0 \pi^0$ contribution was proportional to 
$- {\rm Im} \, \kappa (E)$, where 
$\kappa (E) = (-2 M_{*00} E -i \epsilon )^{1/2}$. This expression
vanishes if $E < 0$. If the $X(3872)$ is a bound state with 
${\rm Re} \, \gamma > 0$, most of the support 
of its line shape is in the region $E<0$. 
Thus the assumption of Ref.~\cite{Hanhart:2007yq} essentially forbids 
the bound state from decaying into $D^0 \bar D^0 \pi^0$. 
This is in direct contradiction to one of the
universal features of an S-wave resonance near threshold. As the
scattering length $a$ increases to $+ \infty$, the binding energy
approaches $1/ (M_{*00}~ a^2)$ and the mean separation of the
constituents approaches $a/2$. In this limit, the decay of the resonance
should be dominated by decays of its constituents. Since $D^0 \pi^0$ is
the largest decay mode of $D^{*0}$, $D^0 \bar D^0\pi^0$ should be a dominant
decay mode of the $X(3872)$ if it is a bound state. 
This flaw in the analysis of Ref.~\cite{Hanhart:2007yq} 
could be removed by using the expression for 
$\kappa (E)$ in Eq.~(\ref{kappa-EGam}) to take into account
the effects of the
decays of the constituents of the $X(3872)$ resonance. 

By comparing Eqs.~(\ref{lsh0:res}) and (\ref{BKres:HKKN}), 
we can express the short-distance factor
$\Gamma_{B^+}^{K^+}$ in terms of the parameters of 
Ref.~\cite{Hanhart:2007yq}:
\begin{equation}
\Gamma_{B^+}^{K^+} = \frac{1}{\pi} \, ({\mathcal B}/g) \, \Gamma [B^+] .
\label{GammaBK:HKKN}
\end{equation}
The best fits in Ref.~\cite{Hanhart:2007yq} give values of 
$\Gamma_{B^+}^{K^+}$ that range from 
$3.8 \times 10^{-13}$ MeV to $4.9 \times 10^{-13}$ MeV. 
This is one order of magnitude larger than the estimate for the
short-distance factor in Eq.~(\ref{GamBKest}) with $\Lambda = m_\pi$.
This requires the square of the 
matrix element $| \mathcal{M} |^2$ for $B^+ \to K^+  + D^{*0} \bar D^0$
near the $D^{*0} \bar D^0$ threshold to be larger than the average value
$\langle | \mathcal{M} |^2 \rangle$ defined by Eq.~(\ref{GamBKD*D})
not only by the threshold resonance factor $m_\pi^2|f(E)|^2$
but also by an additional order of magnitude.

\section{Fits to the energy distributions}
\label{sec:analysis}

The line shapes of $X(3872)$ in the $J/\psi \, \pi^+ \pi^-$
and $D^0 \bar D^0 \pi^0$ decay channels have been measured 
by the Belle and Babar Collaborations for the production
processes $B \to K + X$.  In this section, we fit the 
Belle measurements for the production process
$B^+ \to K^+ + X$ to a theoretical model for the line shapes 
that takes into account the width of the constituent $D^{*0}$
as well as inelastic scattering channels for the charm mesons.

\subsection{Experimental data}

%------------------------------------------------------------------------------------
\begin{table}[tbh]
\begin{tabular}{rrr}
  $E$ ~~~~~ & ~~~$N$~~~ & ~~~$\Delta N$~\\
\hline
$-19.3$ ~~~ &  $-2.10$  &     2.78 \\
$-14.3$ ~~~ &  $-1.10$  &     2.89 \\
 $-9.3$ ~~~ &  $-1.21$  &     3.04 \\
 $-4.3$ ~~~ &    9.60   &     4.83 \\
   0.7  ~~~ &   24.56  &     6.46 \\
   5.7  ~~~ &  $-1.47$  &     2.99 \\
  10.7  ~~~ &  $-1.57$  &     2.99 \\
  15.7  ~~~ &  $-0.58$  &     3.25 \\
\hline
last 7 bins &   28.23  &    10.54
% 20.7 &  0.42 &	3.46 
\end{tabular}
\caption{Belle data on the $ J/\psi \, \pi^+ \pi^-$ energy distribution:
numbers of events $N$ and their uncertainties $\Delta N$
in 5 MeV bins centered at the energies $E$.
The numbers of events $N$ were obtained from Fig.~2b of 
Ref.~\cite{Choi:2003ue} by subtracting the linear experimental background. 
We used only the last 7 data points in our analysis.}
\label{tab:psipipi}
\end{table}
%------------------------------------------------------------------------------------

\begin{figure}[t]
\includegraphics[width=12cm]{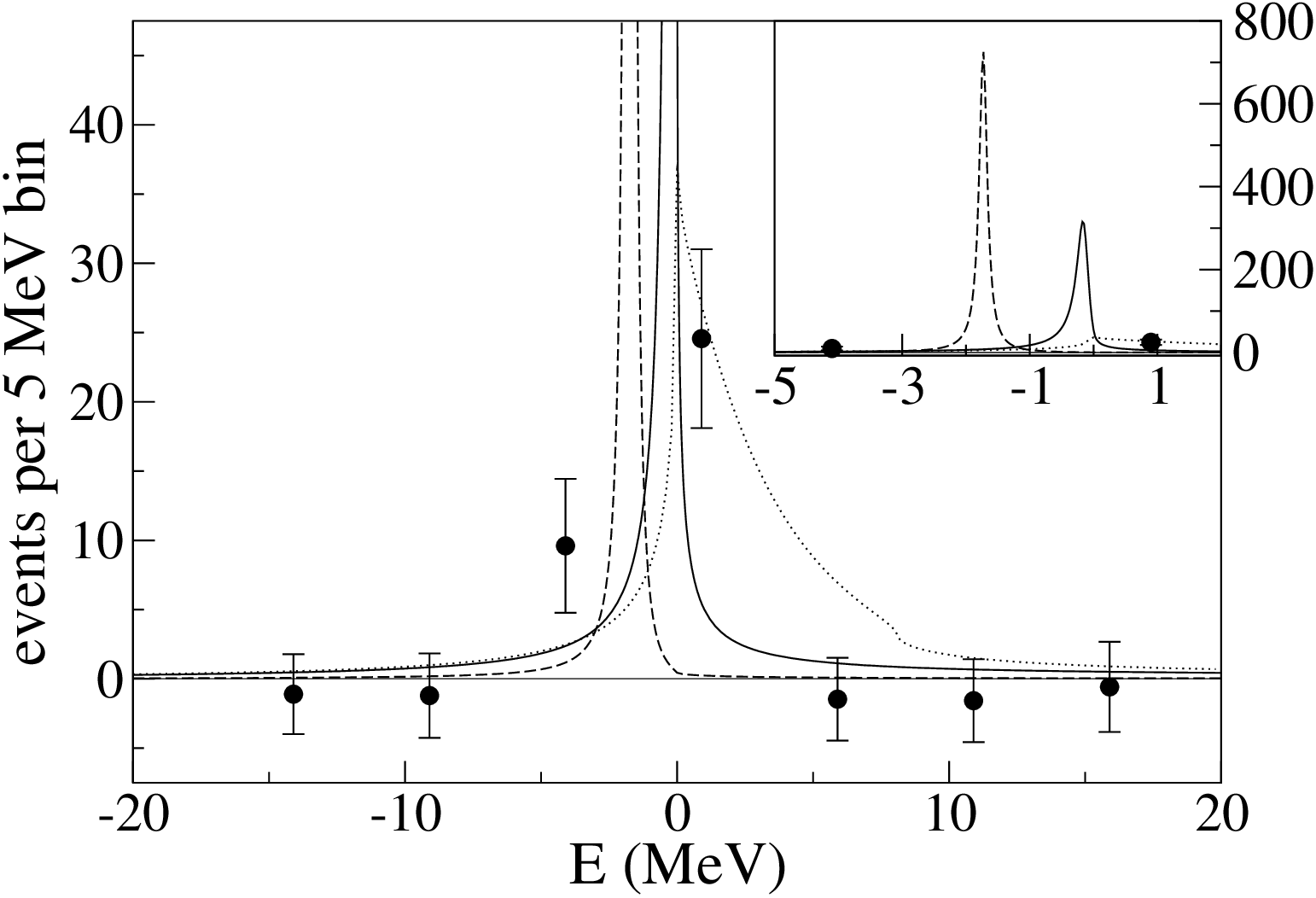}
\caption{
Number of events per 5 MeV bin 
for $B^+ \to K^+ + J/\psi\, \pi^+ \pi^-$
as a function of the total energy $E$ of $J/\psi\, \pi^+ \pi^-$
relative to the $D^{*0} \bar D^0$ threshold.
The data points, which are given in Table~\ref{tab:psipipi},
were obtained by subtracting the linear experimental background
from the data of the Belle collaboration in Ref.~\cite{Choi:2003ue}.
The theoretical curves are the differential number distributions 
$dN/dE$ in Eq.~(\ref{dNdEpsi2pi}) multiplied by the 5 MeV bin width.
The three curves correspond to  
the global minimum of $\chi^2$ (dashed curve), 
the local minimum of $\chi^2$ (solid curve), and the fit 
$A_{\rm Belle}$ of Ref.~\cite{Hanhart:2007yq} (dotted curve).  
The inset shows the peaks of the distributions. 
\label{fig:psipipi} }
\end{figure}

The Belle and Babar Collaborations have both measured the 
energy distribution of $J/\psi \, \pi^+ \pi^-$ near the 
$X(3872)$ resonance for the decay 
$B^+ \to K^+ + J/\psi \, \pi^+ \pi^-$ \cite{Choi:2003ue,Aubert:2004ns}.
The Babar data has larger error bars, so we will consider 
only the Belle data.
The Belle data on the $J/\psi \, \pi^+ \pi^-$ energy distribution
is given in Fig.~2b of Ref.~\cite{Choi:2003ue}.  The figure shows the 
number of events per 
5 MeV bin as a function of $M_{J/\psi \, \pi^+ \pi^-}$ 
from 3820 MeV to 3920 MeV.  If we use the CLEO measurement 
of the $D^0$ mass and the PDG value for the $D^{*0}-D^0$
mass difference, this corresponds to $E$ extending 
from $-51.8$ MeV to $+48.2$ MeV.
In Ref.~\cite{Hanhart:2007yq},  Hanhart et al.~used only the 8 bins 
extending from $-21.8$ MeV to $+18.2$ MeV.  They subtracted the 
linear experimental background shown in Fig.~2b of 
Ref.~\cite{Choi:2003ue} to get the 8 data points given in 
Table~\ref{tab:psipipi}.  Our values for the energies at the
centers of the bins differ from those in Ref.~\cite{Hanhart:2007yq} 
by $-0.2$ MeV, because they used the PDG values for the masses 
$M_0$ and $M_{*0}$ instead of using the CLEO value for $M_0$ 
and the PDG value for $M_{*0} - M_0$.  In our analysis,
we have chosen to omit the first data point in Table~\ref{tab:psipipi}
so that the data points are more symmetric about $E=0$.
We will use a theoretical model for the energy distribution
that is essentially 0 at $E = -19.3$ MeV, so including this point 
would simply increase the $\chi^2$ by a constant 0.57.
The 7 data points used in our analysis are plotted 
in Fig.~\ref{fig:psipipi}.  Note that there are only two bins
in which the data 
differs from 0 by significantly more than one error bar. 

%------------------------------------------------------------------------------------
%------------------------------------------------------------------------------------

%------------------------------------------------------------------------------------
\begin{table}[tbh]
\begin{tabular}{rrr}
     $E$ ~~~~~& ~~~~$N$~~& ~~~$\Delta N$~\\
\hline
  $-5.015$ ~~ &    0.42  &      1.03 \\
  $-0.765$ ~~ &    0.90  &      1.38 \\
    3.485  ~~ &   11.58  &      4.13 \\
    7.735  ~~ &    1.35  &      2.82 \\
   11.985  ~~ &    1.50  &      3.30 \\
   16.235  ~~ &  $-0.89$ &      3.09 \\
\hline
all 6 bins ~~ &   14.86  &      6.96
\end{tabular}
\caption{Belle data on the $D^0 \bar D^0 \pi^0$ energy distribution:
numbers of events $N$ and their uncertainties $\Delta N$
in 4.25 MeV bins centered at the energies $E$.
The numbers of events $N$ were obtained from Ref.~\cite{Majumder}
by subtracting the total experimental background.
Only the last 5 data points were used in the analysis
of Ref.~\cite{Hanhart:2007yq}.}
\label{tab:DDbarpi}
\end{table}
%------------------------------------------------------------------------------------

\begin{figure}[t]
\includegraphics[width=12cm]{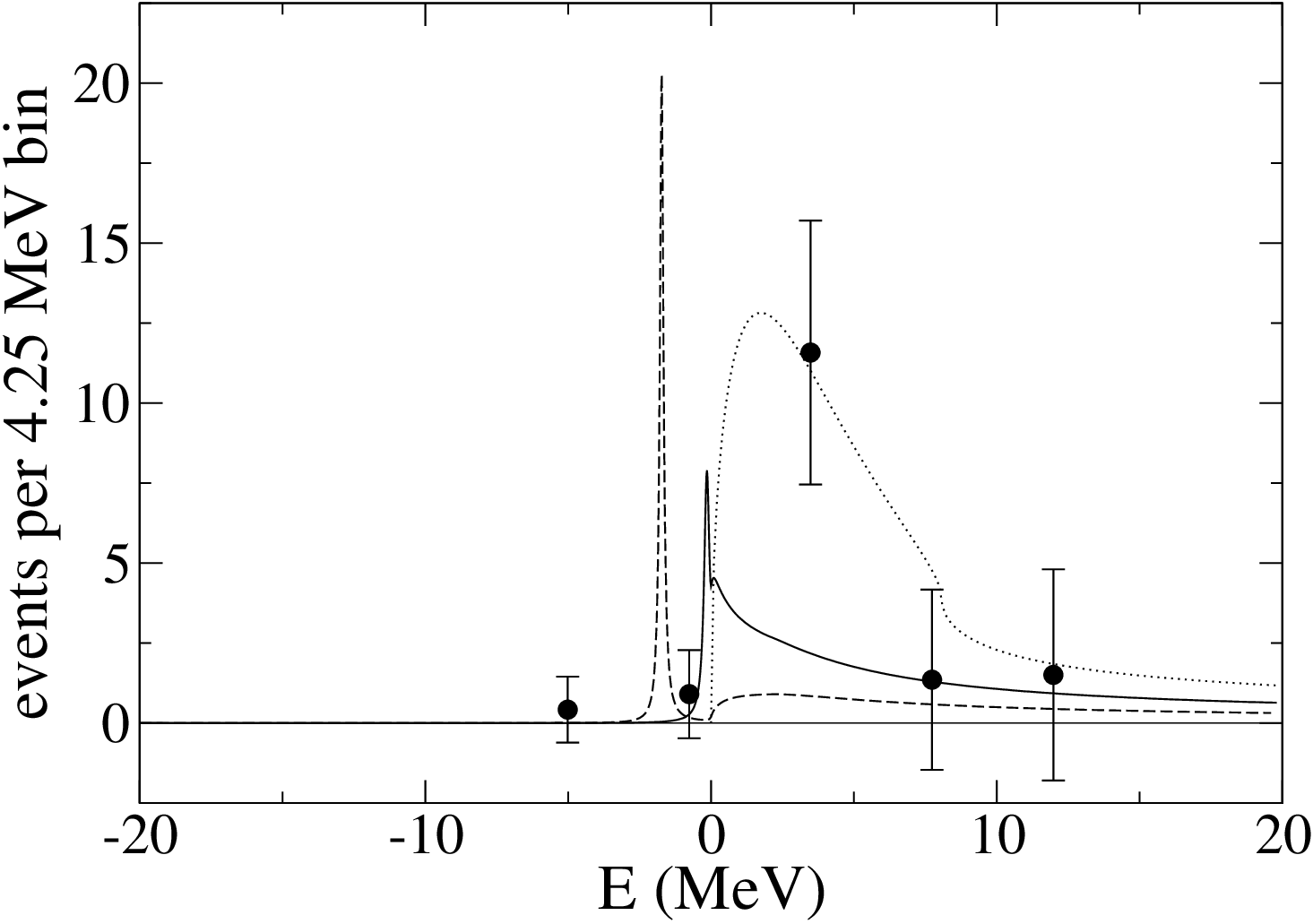}
\caption{
Number of events per 4.25 MeV bin 
for $B^+ \to K^+ + D^0 \bar D^0 \pi^0$
as a function of the total energy $E$ of $D^0 \bar D^0 \pi^0$
relative to the $D^{*0} \bar D^0$ threshold.
The data points, which are given in Table~\ref{tab:DDbarpi},
were obtained by subtracting the total experimental background 
from the data of the Belle Collaboration in Ref.~\cite{Gokhroo:2006bt}.
The theoretical curves are the differential number distributions $dN/dE$
in Eq.~(\ref{dNdEDDbarpi}) multiplied by the 4.25 MeV bin width.
The three curves correspond to 
the global minimum of $\chi^2$ (dashed curve), 
the local minimum of $\chi^2$ (solid curve), and the fit 
$A_{\rm Belle}$ of Ref.~\cite{Hanhart:2007yq} (dotted curve).  
\label{fig:DDbarpi} }
\end{figure}

The Belle Collaboration has measured the 
energy distribution of $J/\psi \, \pi^+ \pi^-$ near the 
$X(3872)$ resonance for the decay 
$B \to K + D^0 \bar D^0 \pi^0$ \cite{Gokhroo:2006bt}.
The $D^0 \bar D^0 \pi^0$ energy distribution is shown in 
Fig.~2a of Ref.~\cite{Gokhroo:2006bt}.  The figure shows the events per 
4.25 MeV bin as a function of 
$M_{D^0 \bar D^0 \pi^0} - 2 M_{D^0} - M_{\pi^0}$ 
from 0 MeV to 76.5 MeV.  This corresponds to $E$ extending from 
$-7.14$ MeV to $+69.36$ MeV.
In Fig.~2a of Ref.~\cite{Gokhroo:2006bt}, the data for 
$B^+ \to K^+ + X$ and $B^0 \to K^0 + X$ are combined in the same plot.
The energy distributions for $B^+$ decay and $B^0$ separately 
were presented at the ICHEP 2006 conference \cite{Majumder}.
In Ref.~\cite{Hanhart:2007yq}, Hanhart et al.~used only the 
data for $B^+ \to K^+ + X$ in 5 bins 
extending from $-2.89$ MeV to $+18.36$ MeV.  They subtracted the 
combinatorial background to obtained the data points for those 5 bins.
To account for the remaining experimental background, which is 
an increasing function of $E$, they added a background term to the 
theoretical expression for $d\Gamma/dE$ and determined its coefficient 
by fitting to the data.  That background term is weakly constrained 
by the 5 data points.  We have therefore chosen to subtract the total 
experimental background instead of only the combinatorial background.
The resulting data points are given in Table~\ref{tab:DDbarpi}.
Our values for the energies at the centers of the bins differ
from those in Ref.~\cite{Hanhart:2007yq} by $-0.34$, because they
used the PDG values for the masses $M_0$ and $M_{*0}$ to determine
the $D^{*0} \bar D^0$ threshold instead of using the CLEO value 
for $M_0$ and the PDG value for $M_{*0} - M_0$.
The first data point in Table~\ref{tab:DDbarpi} was omitted in the 
analysis of Ref.~\cite{Hanhart:2007yq}, 
because it would have given a constant contribution to $\chi^2$ of 0.17. 
In our analysis, we have chosen to include this data point 
even though its effect on our analysis is negligible.  
The 6 data points used in our analysis are plotted 
in Fig.~\ref{fig:DDbarpi}.  Note that there is only one bin in which 
the data differs from 0 by significantly more than one error bar. 

%------------------------------------------------------------------------------------
%------------------------------------------------------------------------------------

The analysis of Ref.~\cite{Hanhart:2007yq} also used the 
Belle measurement of the branching ratio for decays of $X(3872)$ into 
$J/\psi \, \pi^+ \pi^- \pi^0$ and $J/\psi \, \pi^+ \pi^-$ \cite{Abe:2005ix}: 
%-----------------
\begin{equation}
\frac{{\rm Br}[X \to J/\psi \, \pi^+ \pi^- \pi^0]}
{{\rm Br}[X \to J/\psi\, \pi^+ \pi^-]} =
1.0 \pm 0.4 \pm 0.3 .
\label{Br32}
\end{equation}
%-----------------
The signal region for $J/\psi \, \pi^+ \pi^- \pi^0$ included only 
energies $E$ within 16.5 MeV of 3872 MeV. 
Using the CLEO value for $M_0$ and the PDG value for 
$M_{*0} - M_0$, this corresponds to 
$-16.3 \ {\rm MeV} < E < +16.7 \ {\rm MeV}$.

\subsection{Theoretical model}

We summarize our model for the $X(3872)$ lines shapes 
in the channels $J/\psi\, \pi^+ \pi^-$, 
$J/\psi \, \pi^+ \pi^- \pi^0$, and  $D^0 \bar D^0 \pi^0$ channels.
The $J/\psi\, \pi^+ \pi^-$ and $J/\psi \, \pi^+ \pi^- \pi^0$
energy distributions are given by Eq.~(\ref{lsh0:SD}),
while the $D^0 \bar D^0 \pi^0$ energy distribution is given in
Eq.~(\ref{lsh0:DDpi2}).  
The scattering amplitude $f(E)$ is given by Eq.~(\ref{A0-ren})
with ${\rm Im} \, \gamma $  replaced by 
$\Gamma^{J/\psi\, \pi^+ \pi^-}(E) 
+ \Gamma^{J/\psi\, \pi^+ \pi^- \pi^0}(E)$:
%-----------------
\begin{equation}
f(E) = 
\left[ - {\rm Re} \, \gamma 
- i \Gamma^{J/\psi\, \pi^+ \pi^-}(E) 
- i \Gamma^{J/\psi\, \pi^+ \pi^- \pi^0}(E) 
+ \kappa (E) \right]^{-1} .
\label{f-model}
\end{equation}
%-----------------
The energy dependence of the functions 
$\Gamma^{J/\psi\, \pi^+ \pi^-}(E)$ and 
$\Gamma^{J/\psi\, \pi^+ \pi^- \pi^0}(E)$ are shown 
as solid lines in Fig.~\ref{fig:FC}.
Their normalizations are determined by 
$\Gamma^{\psi 2 \pi}_0 \equiv \Gamma^{J/\psi \, \pi^+ \pi^-}(0)$ and 
$\Gamma^{\psi 3 \pi}_0 \equiv \Gamma^{J/\psi \, \pi^+ \pi^- \pi^0}(0)$, 
which we treat as adjustable parameters.
The function $\kappa (E)$ in Eq.~(\ref{f-model}) 
is given in Eq.~(\ref{kappa-EGam}).  The
functions $\Gamma_{*0}(E)$ and ${\rm Br}_{000}(E)$ are given in 
Eqs.~(\ref{GamD*0-E}) and (\ref{Br000}), respectively.
Thus our model has 4 adjustable real parameters:
${\rm Re} \, \gamma$, $\Gamma_{B^+}^{K^+}$,
$\Gamma^{\psi 2 \pi}_0$, and
$\Gamma^{\psi 3 \pi}_0$.

To translate the differential rates $d \Gamma/dE$ into numbers of events
in the Belle experiment, we follow the prescription used in 
Ref.~\cite{Hanhart:2007yq}.  The differential number of 
$J/\psi \, \pi^+ \pi^-$ events at the energy $E$ is
%-----------------
\begin{equation}
\frac{dN}{dE}[J/\psi \, \pi^+ \pi^-] =
\frac{N^{J/\psi \, \pi^+ \pi^-}_{\rm observed} \, \tau[B^+]/\hbar}
    {{\rm Br}[B^+ \to K^+ + X] \; {\rm Br}[X \to J/\psi \, \pi^+ \pi^-]}
\frac{d\Gamma}{dE} [B^+ \to K^+ + J/\psi \, \pi^+ \pi^-] .
\label{dNdEpsi2pi}
\end{equation}
%-----------------
For the number of observed events, we use the central value 
in Ref.~\cite{Choi:2003ue}:
$N^{J/\psi \, \pi^+ \pi^-}_{\rm observed} = 35.7$.
For the product of branching fractions in the denominator, 
we use the central value in Ref.~\cite{Choi:2003ue}, 
which is $1.3 \times 10^{-5}$.
The differential number of $D^0 \bar D^0 \pi^0$
events at the energy $E$ is
%-----------------
\begin{equation}
\frac{dN}{dE}[D^0 \bar D^0 \pi^0] = 
\frac{N^{D^0 \bar D^0 \pi^0}_{\rm observed} \, \tau[B^+]/\hbar}
    {{\rm Br}[B^+ \to K^+ + D^0 \bar D^0 \pi^0]}
    \frac{d\Gamma}{dE} [B^+ \to K^+ + D^0 \bar D^0 \pi^0] .
\label{dNdEDDbarpi}
\end{equation}
%-----------------
For the number of observed events, we use the central value 
in Ref.~\cite{Gokhroo:2006bt}:
$N^{D^0 \bar D^0 \pi^0}_{\rm observed} = 17.4$.
For the branching fraction in the denominator, 
we use the central value in Ref.~\cite{Gokhroo:2006bt}, 
which is $1.02 \times 10^{-4}$.

Following Ref.~\cite{Hanhart:2007yq}, we take into account
the Belle measurement of the branching ratio in Eq.~(\ref{Br32}) 
through a constraint 
on the parameters of our theoretical model.  We demand that 
%-----------------
\begin{equation}
\frac
{\Gamma [B^+ \to K^+ + J/\psi\, \pi^+ \pi^- \pi^0; |E| < 16.5 \ {\rm MeV}]}
{\Gamma [ B^+ \to K^+ + J/\psi\, \pi^+ \pi^-; |E| < 16.5 \ {\rm MeV}]}
= 1.0.
\label{ratiodet}
\end{equation}
%-----------------
This constraint determines the ratio
$\Gamma^{\psi 3 \pi}_0/\Gamma^{\psi 2 \pi}_0$
as a function of the parameters ${\rm Re} \, \gamma$ and 
$\Gamma^{\psi 2 \pi}_0$.
If we combine the errors in Eq.~(\ref{Br32}) in quadrature, 
the error bar on the right side of Eq.~(\ref{ratiodet}) is $\pm 0.5$.
We ignore this error bar and constrain the ratio in Eq.~(\ref{ratiodet}) 
to be 1.0 so that our results can be compared more directly with those 
in Ref.~\cite{Hanhart:2007yq}.%
\footnote{In Ref.~\cite{Hanhart:2007yq}, the partial widths on the 
right side of Eq.~(\ref{ratiodet}) were integrated over the larger 
region $|E| < 20$ MeV, but this difference has a negligible effect 
on the analysis.}

\subsection{Fitting procedure}

%------------------------------------------------------------------------------------
\begin{figure}[t]
\includegraphics[width=12cm]{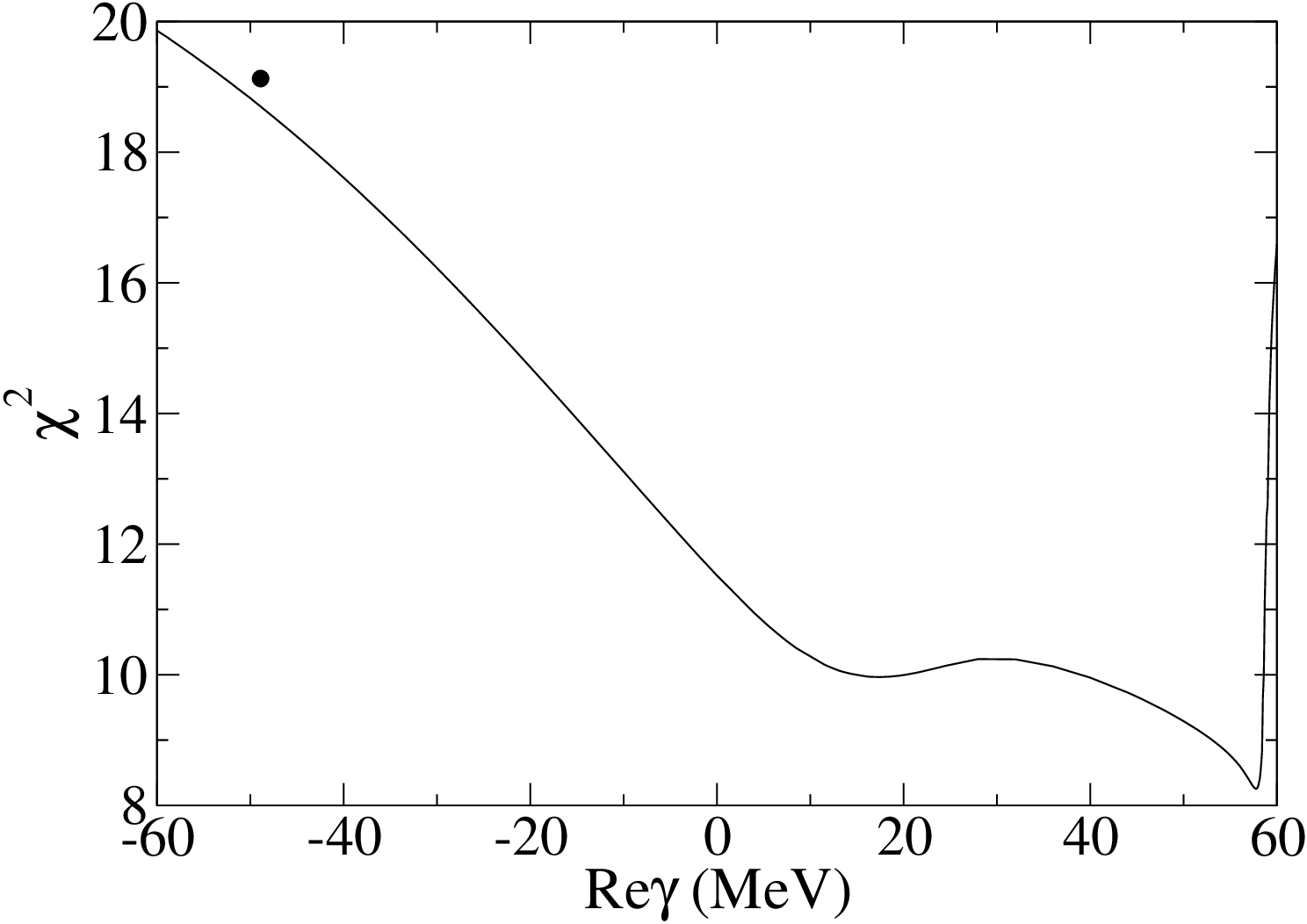}
\caption{
The minimum $\chi^2$ for the 13 data points in 
Figs.~\ref{fig:psipipi} and \ref{fig:DDbarpi}
as a function of ${\rm Re} \, \gamma$.
The $\chi^2$ has been minimized with respect to the parameters 
$\Gamma_{B^+}^{K^+}$, $\Gamma^{\psi 2 \pi}_0$, and
$\Gamma^{\psi 3 \pi}_0$
subject to the constraint in Eq.~(\ref{ratiodet}).
The dot is the $\chi^2$ for the fit $A_{\rm Belle}$
of Ref.~\cite{Hanhart:2007yq}. 
\label{fig:chisq} }
\end{figure}
%------------------------------------------------------------------------------------

One of the most important issues is whether the data on the $X(3872)$ 
is compatible with it being a bound state (corresponding to 
${\rm Re} \, \gamma >0$) or whether it must be a virtual state
(corresponding to ${\rm Re} \, \gamma < 0$) as advocated in 
Ref.~\cite{Hanhart:2007yq}.  We therefore analyze the Belle data 
by fixing the parameter ${\rm Re} \, \gamma$ and minimizing the 
$\chi^2$ with respect to the other 3 adjustable parameters
$\Gamma_{B^+}^{K^+}$, $\Gamma^{\psi 2 \pi}_0$, 
and $\Gamma^{\psi 3 \pi}_0$ subject to the constraint
in Eq.~(\ref{ratiodet}).  The $\chi^2$ is the sum of 13 terms 
corresponding to the last 7 data points in Table~\ref{tab:psipipi}
and the 6 data points in Table~\ref{tab:DDbarpi}.
The constraint in Eq.~(\ref{ratiodet}) determines the ratio 
$\Gamma^{\psi 3 \pi}_0/\Gamma^{\psi 2 \pi}_0$ for fixed
${\rm Re} \, \gamma$.
In Fig.~\ref{fig:chisq}, we show the minimum value of $\chi^2$ 
with respect to variations of the two remaining parameters 
$\Gamma_{B^+}^{K^+}$ and $\Gamma^{\psi 2 \pi}_0$ as a function of 
${\rm Re} \, \gamma$.  The global minimum is $\chi^2 = 8.3$  
at ${\rm Re} \, \gamma = 57.8$ MeV.  There is also a local minimum 
at ${\rm Re} \, \gamma = 17.3$ MeV with $\chi^2 = 10.0$. 
Since ${\rm Re} \, \gamma > 0$ for both the global minimum 
and the local minimum, these fits correspond to a bound state.  
For the sake of comparison, the $A_{\rm Belle}$ fit of 
Ref.~\cite{Hanhart:2007yq} gives $\chi^2 = 19.1$. 
The real part of the inverse scattering length for this fit is
${\rm Re} \, \gamma = -48.9$ MeV.  Since this is negative,
this fit corresponds to a virtual state. 
In Fig.~\ref{fig:chisq}, the value of $\chi^2$ for the 
$A_{\rm Belle}$ fit is shown as a dot that lies just above the line. 

%------------------------------------------------------------------------------------
\begin{table}[t]
\begin{tabular}{l|rcrrr}
     fit       & ~~ Re$\gamma$ ~~ & $\Gamma_{B^+}^{K^+} \times 10^{14}$ & 
~~ $\Gamma^{\psi 2 \pi}_0$ ~ & ~ $\Gamma^{\psi 3 \pi}_0$ ~ & ~~ $\chi^2$ ~~ \\
\hline
global minimum & ~~ $+57.8$ ~~ & 4.1 & ~~ 0.44 ~~ & ~~ 0.50 ~~ & ~~ 8.3 ~~ \\
local minimum  &    $+17.3$ ~~ & 8.3 &    2.3 ~~ &    2.4 ~~ &    10.0 ~~ \\
$A_{\rm Belle}$ of Ref.~\cite{Hanhart:2007yq}   
               &    $-48.9$ ~~ & 38 &    0.93 ~~ &   0.71 ~~ &     19.1 ~~
\end{tabular}
\caption{Parameters (in units of MeV) and values of $\chi^2$ 
for three fits to the subtracted Belle data:
the global minimum of $\chi^2$, 
the local minimum of $\chi^2$, 
and the fit $A_{\rm Belle}$ of Ref.~\cite{Hanhart:2007yq}.}
\label{tab:params}
\end{table}
%------------------------------------------------------------------------------------

In Table~\ref{tab:params}, we list the parameters 
and the values of $\chi^2$ for the global minimum, the local minimum, 
and the fit $A_{\rm Belle}$ of Ref.~\cite{Hanhart:2007yq}.
For the fit $A_{\rm Belle}$, ${\rm Re} \gamma$ is the real part 
of the inverse scattering length, $\Gamma_{B^+}^{K^+}$ is given 
in Eq.~(\ref{GammaBK:HKKN}), and $\Gamma^{\psi 2 \pi}_0$ and
$\Gamma^{\psi 3 \pi}_0$ are the values of 
$\Gamma_{\pi^+\pi^-J/\psi}(E)/g$ and
$\Gamma_{\pi^+\pi^-\pi^0J/\psi}(E) /g$ at $E=0$.  Our order-of-magnitude 
estimate for $\Gamma_{B^+}^{K^+}$ is given in Eq.~(\ref{GamBKest})
with $\Lambda = m_\pi$. The values of $\Gamma_{B^+}^{K^+}$
for the global minimum and the local minimum of $\chi^2$
are more consistent with this estimate than the fit  
$A_{\rm Belle}$ of Ref.~\cite{Hanhart:2007yq}. 

The line shapes of $X(3872)$ in the $J/\psi \, \pi^+ \pi^-$
and $D^0 \bar D^0 \pi^0$ decay channels for various fits are shown 
together with the Belle data in
Figs.~\ref{fig:psipipi} and \ref{fig:DDbarpi}, respectively.
The line shapes corresponding to the local minimum, the global minimum, 
and the fit $A_{\rm Belle}$ of Ref.~\cite{Hanhart:2007yq} are shown 
as solid, dashed, and dotted lines, respectively.  The data points 
in these figures should be compared to the average values of the curves 
over the appropriate bins centered on the data points. 
The global minimum of $\chi^2$ is somewhat pathological in that 
for both $J/\psi \, \pi^+ \pi^-$ and $D^0 \bar D^0 \pi^0$ 
the integral of the line shape over a bin is largest 
not in the bin with the highest data point, 
but in the next lower bin.
For the local minimum of $\chi^2$, the integrated line shape 
gives a good fit to the data point in the highest bin for 
$J/\psi \, \pi^+ \pi^-$ but it is smaller than 
the data point in the highest bin for 
$D^0 \bar D^0 \pi^0$ by more than two standard deviations.

\section{Discussion}
\label{sec:disc}

We have developed an approximation to the line shapes of the $X(3872)$ 
that takes into account the width of its constituents $D^{*0}$ or 
$\bar D^{*0}$ as well as inelastic scattering channels for 
$D^{*0} \bar D^0$ and $D^0 \bar D^{*0}$.  The best combined fit 
to the Belle data on the energy distributions for 
the $X$ resonance in the 
$J/\psi \, \pi^+ \pi^-$ and $D^0 \bar D^0 \pi^0$ channels 
corresponds to a bound state with ${\rm Re} \, \gamma >0$, 
although a virtual state with ${\rm Re} \, \gamma < 0$
is not excluded.  Our results are in contradiction to the 
conclusions of Ref.~\cite{Hanhart:2007yq}.  By fitting essentially 
the same data, they concluded that the $X(3872)$ must be a virtual state.
The flaw in their analysis was that they did not allow for decays 
of the constituent $D^{*0}$ or $\bar D^{*0}$ in the case where
the $X$ is a bound state.

Our theoretical model for the line shapes is based on the 
assumption that the inverse scattering length in the $(D^* \bar D)_+^0$
channel is small compared to all other relevant momentum scales.
The line shapes are determined by the scattering amplitude $f(E)$ 
for charm mesons in the $(D^* \bar D)_+^0$ channel given in
Eq.~(\ref{f-model}).  This amplitude contains no information about 
some of the other nearby thresholds, including the $D^{*+} D^-$ 
threshold at $E = +8.1$ MeV and the $D^0 \bar D^0 \pi^0$ 
threshold at $E = -7.1$ MeV.  Thus the line shapes can be expected 
to be accurate only when $|E|$ is much less than 7 MeV.
However the maxima of the line shapes are well within
the region of validity.  For the local minimum of $\chi^2$, 
the peaks in the line shapes are near 
$E = -0.17$ MeV for $J/\psi \, \pi^+ \pi^-$ and near 
$E = -0.14$ MeV for $D^0 \bar D^0 \pi^0$.
For the global minimum of $\chi^2$,
the peaks in the line shapes are near 
$E = -1.7$ MeV for both $J/\psi \, \pi^+ \pi^-$
and $D^0 \bar D^0 \pi^0$.
The effects of the other thresholds would be dramatic only when
$|E|$ is large enough that the line shapes are already small.
They would be unlikely to change our qualitative conclusion 
that the best fit to the data corresponds to a bound state.

The range of validity of our model for the line shapes 
could be extended by taking into account explicitly scattering 
channels for the charged charm mesons $D^{*+} D^-$ and $D^+ D^{*-}$.
A first step in this direction has been taken by Voloshin
\cite{Voloshin:2007hh}.  He showed that the isospin symmetry 
breaking pattern of QCD provides interesting constraints 
on the line shapes.  In particular, he predicted a zero in 
the line shape of $X(3872)$ in the $J/\psi \, \pi^+ \pi^- \pi^0$
channel between the $D^{*0} \bar D^0$ and $D^{*+} D^-$ thresholds.
We have carried out a more rigorous two-channel analysis
\cite{Braaten-Lu}.  Given a plausible dynamical assumption,
we find that the line shape of $X$
in the $J/\psi \, \pi^+ \pi^- \pi^0$ channel has a zero
between the $D^{*0} \bar D^0$ and $D^{*+} D^-$ thresholds
in the decays $B^+ \to K^+ + X$ but not in the decays 
$B^0 \to K^0 + X$. 
We also find that the line shape of $X$ in the 
$J/\psi \, \pi^+ \pi^-$ channel can have a zero 
below the $D^{*0} \bar D^0$ threshold in the decays 
$B^0 \to K^0 + X$.

The accuracy of our predictions for the line shapes could be 
further improved by taking into account pions explicitly.
The system consisting of $D^{*0} \bar D^0$,
$D^0 \bar D^{*0}$, and $D^0 \bar D^0 \pi^0$ states with energies
near the $D^0 \bar D^{*0}$ threshold can be described by a 
nonrelativistic effective field theory.  The simplest such theory 
has S-wave scattering in the $(D^* \bar D)_+^0$ channel
and $\pi^0$ couplings that allow the decay $D^{*0} \to D^0 \pi^0$.
Fleming, Kusunoki, Mehen, and van Kolck developed power-counting 
rules for this effective field theory and showed that the pion
couplings can be treated perturbatively \cite{Fleming:2007rp}.
They used the effective field theory to calculate the decay rate for
$X(3872) \to D^0 \bar D^0 \pi^0$ to next-to-leading order in the 
pion coupling.

In applying this effective field theory to the line shapes of the 
$X(3872)$, one complication that will be encountered is infrared 
singularities at the $D^{*0} \bar D^0$ threshold that are related 
to the decay $D^{*0} \to D^0 \pi^0$.  This problem has been 
analyzed in a simpler model with spin-0 particles and 
momentum-independent interactions \cite{Braaten:2007ct}.
The problem was solved by a resummation of perturbation theory
that takes into account the perturbative shift of the 
$D^0 \bar D^{*0}$ threshold into the complex energy plane
because of the nonzero width of the $D^{*0}$.

In summary, the establishment of the quantum numbers of the 
$X(3872)$ as $1^{++}$ and the measurement of its mass imply 
that it is either a charm meson molecule or a charm meson 
virtual state.  The existing data favor a charm meson molecule,
but a virtual state is not excluded.
To decide conclusively between these two possibilities
will require more extensive data on the line 
shapes of the $X(3872)$ in various decay channels 
and for various production processes.

\begin{acknowledgments}
This research was supported in part by the Department of Energy
under grant DE-FG02-91-ER40690.  We thank the authors
of Ref.~\cite{Hanhart:2007yq} for sharing with us the data points 
used in their analysis.
\end{acknowledgments}

%\newpage


\end{document}